\documentclass{article}

\usepackage{PRIMEarxiv}
\usepackage{siunitx}
\usepackage[utf8]{inputenc} 
\usepackage[T1]{fontenc}    
\usepackage{hyperref}       
\usepackage{url}            
\usepackage{booktabs}       
\usepackage{amsfonts}       
\usepackage{nicefrac}       
\usepackage{microtype}      
\usepackage{lipsum}
\usepackage{fancyhdr}       
\usepackage{graphicx}       
\graphicspath{{media/}}     
\usepackage{lineno}
\usepackage{lscape}    
\usepackage{longtable}

\title{Artificial Intelligence for Microbiology and Microbiome Research
}

\author{
  Xu-Wen Wang, Tong Wang \\
  Channing Division of Network Medicine \\
  Department of Medicine \\
  Brigham and Women's Hospital and Harvard Medical School \\
  Boston, MA 02115 USA\\
    \And
  Tong Wang \\
  Channing Division of Network Medicine \\
  Department of Medicine \\
  Brigham and Women's Hospital and Harvard Medical School \\
  Boston, MA 02115 USA\\
  Department of Biological Sciences \\
  Purdue University \\
  West Lafayette, IN 47907, USA \\
    \And
  Yang-Yu Liu \\
  Channing Division of Network Medicine \\
  Department of Medicine \\
  Brigham and Women's Hospital and Harvard Medical School \\
  Boston, MA 02115 USA\\
  \texttt{yyl@channing.harvard.edu} \\
}

\begin{document}
    \maketitle

\begin{abstract}
Advancements in artificial intelligence (AI) have transformed many scientific fields, with microbiology and microbiome research now experiencing significant breakthroughs through machine learning applications. This review provides a comprehensive overview of AI-driven approaches tailored for microbiology and microbiome studies, emphasizing both technical advancements and biological insights. We begin with an introduction to foundational AI techniques, including primary machine learning paradigms and various deep learning architectures, and offer guidance on choosing between traditional machine learning and sophisticated deep learning methods based on specific research goals. The primary section on application scenarios spans diverse research areas, from taxonomic profiling, functional annotation \& prediction, microbe-X interactions, microbial ecology, metabolic modeling, precision nutrition, clinical microbiology, to prevention \& therapeutics. Finally, we discuss challenges in this field and highlight some recent breakthroughs. Together, this review underscores AI's transformative role in microbiology and microbiome research, paving the way for innovative methodologies and applications that enhance our understanding of microbial life and its impact on our planet and our health.
\end{abstract}

\tableofcontents

\section{Introduction}\label{INTRODUCTION}
For over 3.5 billion years, our planet and its inhabitants have been shaped by various microorganisms~\cite{blaser2016toward}. For example, Cyanobacteria, through photosynthesis, produced oxygen and contributed to the Great Oxygenation Event around 2.4 billion years ago, making the Earth hospitable for aerobic life~\cite{lyons2014rise}. Certain bacteria, like Rhizobium, fix atmospheric nitrogen into forms usable by plants, supporting plant growth and agriculture~\cite{oldroyd2014biotechnological}. Commensal microbes in human and animal guts aid in digestion and nutrient absorption, essential for health and survival~\cite{turnbaugh2007human}. Similarly, some microbes can break down organic matter, recycling nutrients in ecosystems, which is vital for maintaining soil fertility and ecosystem balance~\cite{gadd2010metals}. Given the profound influence microorganisms have had on the evolution of life and the functioning of ecosystems, advancing microbiology research is crucial for understanding and harnessing these processes to benefit health, agriculture, and environmental sustainability.

It is not a big surprise that disrupted microbial communities (or microbiomes) can have a huge impact on our planet and ourselves. Indeed, agricultural practices, such as excessive use of chemical fertilizers and pesticides, can disrupt soil microbiomes, leading to reduced soil fertility and increased vulnerability to erosion~\cite{geisseler2014long}. Runoff containing pollutants and antibiotics can significantly disrupt the microbiomes of freshwater and marine ecosystems, leading to changes in water quality and impacting the health of aquatic life by altering the natural balance of microbial communities within the environment; this can potentially promote the growth of harmful bacteria and disrupt critical ecological processes like nutrient cycling~\cite{grenni2018ecological,martinez2009environmental}. Many human diseases have been associated with disrupted microbiomes, including acne, eczema, dental caries, obesity, malnutrition, inflammatory bowel disease, asthma/allergies, hardening of arteries, colorectal cancer, type 2 diabetes, as well as neurological conditions such as autism, anxiety, depression, and post-traumatic stress disorder, etc~\cite{young2017role,afzaal2022human,ke2023association}. Gaining a deeper understanding of the activities of microbial communities, both within and around us, can greatly benefit our health and the health of our planet. This explains why in the past decades the microbiome has been a very active research topic in microbiology. 

Artificial Intelligence (AI) focuses on creating intelligent machines that can execute tasks that usually need human intelligence. AI emerged as an academic discipline at the 1956 Dartmouth conference, shaped by pioneering work by Warren McCulloch, Walter Pitts, and Alan Turing on neural networks and machine intelligence.  At first, AI research concentrated on symbolic reasoning, including early applications in biomedicine, such as the MYCIN expert system for diagnosing bacterial infections. Meanwhile, machine learning developed, showcasing algorithms that improved through data training. Despite early excitement and positive forecasts, the pace of AI advancement decelerated over the following decades, hindered by hardware constraints and unmet expectations, leading to a period known as ``AI winter." However, the domain continued to progress, incorporating probabilistic methods to manage uncertainty. In around 2010, a new phase in AI emerged, fueled by breakthroughs in deep learning frameworks, the advent of powerful hardware (e.g., GPUs), open-source software tools, and greater access to extensive datasets (e.g., ImageNet~\cite{5206848}). In 2012, significant breakthroughs occurred when AlexNet (a deep learning architecture based on the convolutional neural network) surpassed preceding machine learning methodologies in visual recognition~\cite{krizhevsky2012imagenet}. The subsequent innovations, particularly the Transformer (a deep learning architecture initially developed for machine translation) introduced in 2017~\cite{vaswani2017attention}, triggered an ``AI boom" marked by considerable investment. This surge in investment led to a wide range of AI applications by the 2020s, accompanied by increasing concerns regarding its societal implications and the pressing need for regulatory measures. 

Traditional microbiologists excel in image analysis skills for identifying pathogens in Gram stains, ova and parasite preparations, blood smears, and histopathologic slides. They also characterize colony growth on agar plates to identify microorganisms and assess their properties, such as antibiotic susceptibility or ecological roles, often complementing these techniques with modern molecular methods. AI advances in computer vision can automate these processes, supporting timely and accurate diagnoses~\cite{peiffer2020machine,burns2023use}. Advances in sequencing technologies, especially next-generation sequencing, enable substantial numbers of samples to be processed rapidly and cost-efficiently~\cite{cox2013sequencing}. The accessibility of large-scale microbiome datasets propelled the development of numerous AI (especially machine learning) approaches in microbiome studies~\cite{marcos2021applications,wu2021towards,qu2019application, ghannam2021machine,cammarota2020gut,moreno2021statistical,namkung2020machine,li2022machine,yecsilyurt2022microbiome,metcalf2017microbiome, goodswen2021machine,roy2024deep,gerber2024ai,lim2022artificial,zhu2020application,zeng_applying_2021,mccoubrey_harnessing_2021,loganathan_influence_2022,knights_supervised_2011,hernandez_medina_machine_2022,asnicar2023machine,malakar2024understanding,lin2021artificial,jiang2022machine,iadanza2020gut,roy2024deep,tonkovic2020literature,loganathan2022influence,mathieu2022machine,krause2021analyzing,abavisani2024deciphering,he2022advances,kumar2022artificial,wu2024multi,roy2024deep,yan2025recent}. However, a comprehensive review of existing applications of AI techniques in microbiology and microbiome research is still lacking. This review article aims to fill this gap. The following sections are organized as follows. We first briefly describe various AI subfields, focusing on machine learning and the three basic machine learning paradigms. Next, we elaborate on the different deep learning techniques categorized according to their underlying inductive biases and learning objectives. Then, we systematically review the various applications of AI techniques in microbiology and microbiome research. Finally, we discuss outstanding challenges and future directions of AI for microbiology and microbiome research.

\phantomsection
\section{Artificial Intelligence Techniques}
The multiple subfields of AI research are focused on specific objectives and the utilization of distinct tools. The conventional objectives of AI research encompass searching, knowledge representation, reasoning, planning, learning, communicating, perceiving, and acting~\cite{russell2021artificial}. Most AI applications in microbiology and microbiome research rely on machine learning, which is the focus AI subfield of this Review. 

\subsection{Machine Learning Paradigms}
Machine learning is a subfield of AI that employs algorithms and statistical models, enabling machines to learn from data and improve their performance on specific tasks over time~\cite{10.5555/1162264}. Machine learning is typically categorized into three primary learning paradigms: {\bf supervised learning}, {\bf unsupervised learning}, and {\bf reinforcement learning}. These paradigms differ in the specific tasks they can address as well as in the manner in which data is presented to the computer. Generally, the nature of the task and the data directly influence the selection of the appropriate paradigm.  

Supervised learning involves using labeled datasets, where each data point is linked to a class label. The algorithms in this approach aim to create a mathematical function that connects input features to the expected output values, relying on these labeled instances. Common uses include classification and regression. Traditional machine learning methods for classification/regression include Logistic Regression, Na\"ive Bayes, Support Vector Machine (SVM), Random Forest (RF), Extreme Gradient Boosting (XGBoost), etc. Those methods have been heavily used in microbiology and microbiome research. 

In unsupervised learning, algorithms analyze unlabeled data to detect patterns and relationships without any defined categories. This process uncovers similarities in the dataset and includes techniques like clustering, dimensionality reduction, and association rules mining. Classical unsupervised learning methods include k-means clustering, Principal Component Analysis (PCA), Principal Coordinate Analysis (PCoA), t-distributed stochastic neighbor embedding (t-SNE), and Uniform Manifold Approximation and Projection (UMAP) for dimension reduction, and the Apriori algorithm for association rules mining. Among them, PCoA is a commonly used tool in microbiome data analysis, particularly valuable for visualizing and interpreting the differences in compositions of different samples.  

Reinforcement learning (RL) focuses on enabling intelligent agents to learn through trial-and-error in a dynamic environment to maximize their cumulative rewards~\cite{sutton1998reinforcement}. Without labeled datasets, these agents make decisions to maximize rewards, engaging in autonomous exploration and knowledge acquisition, which is crucial for tasks that are difficult to program explicitly. RL has seen limited adoption in microbiology and microbiome research to date, despite its proven efficacy in other domains like robotics, game playing, and drug optimization. This scarcity stems from challenges inherent to microbial systems—such as high-dimensional data from metagenomics, complex community dynamics, and the need for interpretable models in biological contexts—making RL's trial-and-error paradigm harder to adapt than supervised learning approaches like classification or clustering, which dominate current microbiome studies.

Integrating these paradigms can often lead to better outcomes. For instance, {\bf semi-super-vised learning} finds a middle ground by utilizing a small set of labeled data alongside a larger collection of unlabeled data. This method harnesses the strengths of both supervised and unsupervised learning, making it a cost-effective and efficient way to train models when labeled data is scarce. In situations where obtaining high-quality labeled data is difficult, {\bf self-supervised learning} presents a viable alternative~\cite{geiping2023cookbook}. In this framework, models are pre-trained on unlabeled data, with labels generated automatically in subsequent iterations. Self-supervised learning effectively converts unsupervised machine learning challenges into supervised tasks, improving learning efficiency. 

{\bf Transfer learning} is another powerful machine learning technique, which involves taking a pre-trained model on a large dataset and fine-tuning it on a smaller, task-specific dataset~\cite{pan2009survey,tan2018survey}. This approach leverages the knowledge acquired by the model during pre-training to improve performance on a new task. Transfer learning can be applied within both supervised and unsupervised learning paradigms, meaning it can utilize knowledge learned from either labeled or unlabeled data depending on the situation; essentially, transfer learning "transfers" the learned representations from one task to another, regardless of whether the original task was supervised or unsupervised.

Note that both self-supervised learning and transfer learning leverage {\bf pre-trained models} to improve performance on new tasks, but the key difference is that self-supervised learning generates its own labels, often called ``pseudo-labels”, from unlabeled data during the pre-training phase, while transfer learning relies on existing labeled or unlabeled data for pre-training. Both self-supervised learning and transfer learning are extensively used in the training of {\bf large language models} (LLMs), with self-supervised learning often being the primary method for pre-training on massive amounts of unlabeled data, while transfer learning allows the pre-trained model to be adapted to specific downstream tasks with fine-tuning on smaller labeled datasets. LLMs tailored for biology, e.g., genomic and protein language models~\cite{ji_dnabert_2021, Hwang2024, rives2021biological, elnaggar2021prottrans, brandes2022proteinbert,nguyen2024sequence,merchant2024semantic,brixi2025genome}, have numerous applications in microbiology and microbiome research. These models, trained on vast amounts of biological sequence data, can generate insights and predictions that are valuable across various areas in microbiology and microbiome research. 

\phantomsection
\subsection{Deep learning architectures}
As a subfield of machine learning, deep learning represents a further specialization that utilizes deep neural networks to process and analyze large datasets, allowing for the automatic identification of patterns and the solving of complex problems. The reason why we often need a deeper rather than a wider neural network is that, if we regard a neural network as a function approximator, the complexity of the approximation function will typically grow exponentially with depth (not width). In other words, with the same number of parameters, a deep and narrow network has stronger expressive power than a shallow and wide network~\cite{montufar2014number,pascanu2014construct,raghu2017expressive,serra2018bounding,arora2018understanding}.


Deep learning architectures can be broadly categorized according to their underlying \textbf{inductive biases}—the structural assumptions that guide how they process data. Models designed for tabular and dense features often rely on multilayer perceptrons (MLPs) and increasingly incorporate specialized transformer-based architectures such as TabTransformer~\cite{huang2020tabtransformer}, TabNet~\cite{arik2021tabnet}, and FT-Transformer~\cite{gorishniy2021revisiting} to better capture feature interactions. Convolutional neural networks (CNNs), e.g., LeNet-5~\cite{lecun2002gradient}, AlexNet~\cite{krizhevsky2012imagenet}, ResNet~\cite{he2016deep}, U-Net~\cite{ronneberger2015u}, ConvNeXt~\cite{liu2022convnet}, introduce locality and translation-equivariance priors, making them particularly effective for images and other spatially structured inputs. Attention-based and Transformer architectures, e.g., Transformer~\cite{vaswani2017attention}, GPT~\cite{radford2018improving}, BERT~\cite{devlin2019bert}, ViT~\cite{dosovitskiy2021an}, Perceiver~\cite{jaegle2021perceiver}, Llama~\cite{touvron2023llama}, emphasize flexible, content-dependent interactions and now dominate natural language processing and are increasingly applied to vision and multimodal learning. For time-series and sequential data, recurrent neural networks (RNNs)~\cite{rumelhart1985learning,jordan1997serial} and its variants: Long short-term memory (LSTM)~\cite{hochreiter1997long} and Gated Recurrent Unit (GRU)~\cite{cho2014properties} have historically been used. Receptance Weighted Key Value (RWKV)~\cite{peng-etal-2023-rwkv} is a hybrid model architecture that combines the efficient parallelizable training of Transformers with the efficient inference of RNNs. Originated in control systems engineering, state-space models (SSM), e.g., High-order Polynomial Projection Operators (HiPPO)~\cite{gu2020hippo}, structured state space sequence (S4) model~\cite{gu2022efficiently} and Mamba~\cite{gu2024mamba}, offer efficient long-range sequence modeling through continuous-time and linear-state dynamics. They have proven to match the performance capacity of transformer-based models while offering superior speed and efficiency. Graph neural networks (GNNs) and its variants e.g., graph convolutional network (GCN)~\cite{kipf2016semi}, graph attention network (GAT)~\cite{velivckovic2018graph}, graph recurrent network (GRN)~\cite{seo2018structured}, incorporate relational inductive biases tailored to graph-structured inputs such as molecular graphs. By effectively encoding the structural information of a graph into the standard Transformer architecture, Graphormer~\cite{ying2021transformers} has attained excellent results on a broad range of graph representation learning tasks.  Finally, neural operator models, e.g., Deep Operator Networks (DeepONets)~\cite{lu2021learning}, Fourier Neural Operator (FNO)~\cite{li2021fourier}, Laplace Neural Operator~\cite{cao2024laplace}, extend deep learning to learn solution operators for differential equations, enabling data-driven modeling across scientific and engineering domains. Together, these architectural families illustrate how inductive biases shape modern deep learning systems to match the structural properties of diverse data modalities.

Deep learning models are fundamentally distinguished by their \textbf{learning objectives}, which define the mathematical target optimized during training. Discriminative objectives model the conditional probability $p(y|x)$ to perform classification or regression, exemplified by ResNet~\cite{he2016deep} for image classification and BERT~\cite{devlin2019bert} (with a classification head) for text categorization. Generative objectives learn the data distribution $p(x)$ or $p(x,y)$, split into explicit density models (e.g., Variational Autoencoders (VAEs)~\cite{kingma2013auto}, Pixel Recurrent Neural Networks (PixelRNN)~\cite{van2016pixel}, and Normalizing Flows (NF)~\cite{kingma2018glow}) which provide tractable likelihoods) and implicit sampling models (e.g., Generative Adversarial Networks (GANs)~\cite{goodfellow2016nips} and Diffusion Models (including Denoising Diffusion Probabilistic Models (DDPMs)~\cite{ho2020denoising}, Stable Diffusion~\cite{podell2024sdxl} and Sora~\cite{DBLP:journals/corr/abs-2402-17177}) which generate high-fidelity samples without explicit densities. Contrastive/Self-Supervised objectives learn invariant representations by maximizing similarity between augmented views of the same instance, as in SimCLR~\cite{chen2020simple}, BYOL~\cite{grill2020bootstrap}, and DINO~\cite{caron2021emerging} for images or SimCSE~\cite{gao-etal-2021-simcse} for text. Reconstructive objectives minimize input-output divergence, typically via autoencoding, with Masked Autoencoders (MAE) and standard Autoencoders (AE) learning compressed latent representations. Predictive objectives forecast future tokens or frames in a sequence, powering large language models like GPT and LLaMA (next-token prediction) and video generation models like Phenaki~\cite{villegas2023phenaki}. These objectives are largely orthogonal to architecture and supervision, enabling flexible combinations in modern foundation models.

\phantomsection
\subsection{Traditional Machine Learning vs. Deep Learning: When to Use Each}
As we will see later, many deep learning architectures (e.g., MLP, CNN, RNN, Transformer, GAN, AE), their variants (e.g., LSTM, VAE, GCN, BERT), their hybrids (e.g., CNN+LSTM, AE+GAN; GAN+CNN, AE+CNN), and their hybrids with traditional machine learning models (e.g., AE+SVM, CNN+Random Forest) have been widely used in microbiology and microbiome research.  However, we have to emphasize that we do not always need advanced deep learning techniques for microbiology and microbiome research. Sometimes we do not need deep learning at all. Logistic Regression or Random Forest might work very well. Choosing between deep learning and traditional machine learning methods depends on data characteristics, the specific problem at hand, available computational resources, and the need for model interpretability. Traditional methods are generally preferred for smaller, structured datasets and scenarios requiring interpretability (such as clinical applications), while deep learning excels with large, unstructured datasets and complex tasks requiring high performance.

If we decide to apply or develop deep learning methods to solve our problem, there is a systematic approach~\cite{goodfellow2016deep}. First, we need to choose performance metrics tailored to the task and dataset characteristics, such as Silhouette Index, Adjusted Rand Index, or Normalized Mutual Information for clustering; FID or Inception Score for generative models; accuracy, precision, recall, F1-score, AUC-ROC, or AUC-PR for classification; MSE, MAE, RMSE, or R² for regression; and MASE or sMAPE for time series forecasting, ensuring alignment with the problem's objectives and constraints. Second, we need to find the default baseline deep learning models based on the data structure. For supervised learning tasks that involve fixed-size vector inputs, it is advisable to utilize a feedforward network featuring fully connected layers (e.g., MLP) or transformer-based architectures (e.g., TabTransformer, TabNet, or FT-Transformer). If the input possesses a known topological structure, such as images or graphs, opting for CNN or its variants (e.g.,GCN) is recommended. CNNs excel in processing Euclidean data, like microbial imaging or spatial microbiome distributions, by leveraging local receptive fields and weight sharing, whereas GCNs are suited for non-Euclidean data, such as microbiome interaction graphs or phylogenetic trees, by aggregating relational information across nodes to model microbial community dynamics or taxonomic dependencies. When dealing with inputs or outputs that form sequences, we should consider using RNN and its variant (e.g., LSTM or GRU) or Transformer. 1D CNN or temporal convolutional network (TCN) might also work. Depending on the task, a hybrid deep learning model could also be considered. Third, we need to establish a reasonable end-to-end system, which involves choosing the appropriate optimization algorithm (e.g., SGD with momentum, Adam) and incorporating regularization (via early stop, dropout, or batch normalization). Finally, we need to measure the performance and determine how to improve it. We can either gather more training data or tune hyperparameters (e.g., learning rate, number of hidden units) via grid search or random search. 

\phantomsection
\subsection{Data Types in AI Applications for Microbiome Research}
At the intersection of AI and microbiology/microbiome research, the choice of data type is critical for uncovering microbial functions, interactions, and ecological dynamics. AI models—from classical machine learning to deep learning—leverage diverse omics layers to predict functions, classify taxa, and simulate communities. Core data types include metagenomics (DNA sequences of communities), metatranscriptomics (RNA expression), metaproteomics (protein profiles), metabolomics (microbial metabolites), and meta-epigenomics (epigenetic modifications). Together, these offer a multidimensional view that enables AI to detect complex relationships often missed by traditional analyses.

Equally important are the input formats. Raw sequencing data (e.g., FASTQ) are information-rich but computationally demanding. To reduce dimensionality, they can be transformed into k-mers or relative abundances from profiling tools, which integrate more easily with standard models but risk losing biological nuance. More advanced approaches use LLM embeddings (e.g., ESM-2~\cite{lin2023evolutionary}, ProtGPT2~\cite{ferruz2022protgpt2}, ProtBERT~\cite{brandes2022proteinbert}), which encode sequences into dense vectors capturing higher-order relationships and enabling transfer learning.

The choice of data type and format shapes model performance and generalizability. Metagenomics offers richness but risks overfitting without normalization or augmentation. Metaproteomics and metabolomics add functional depth but suffer from missing values and batch effects unless carefully preprocessed. Raw sequences require heavy computation and advanced architectures, while relative abundances trade scalability for detail. LLM embeddings improve transferability across datasets but demand careful fine-tuning to avoid bias. Ultimately, hybrid, multi-omics integration—when harmonized properly—yields more robust and generalizable AI models, though it requires strong interdisciplinary expertise.

\section{Application Scenarios}
\label{sec:3}
\begin{figure*}
  \includegraphics[width=\linewidth]{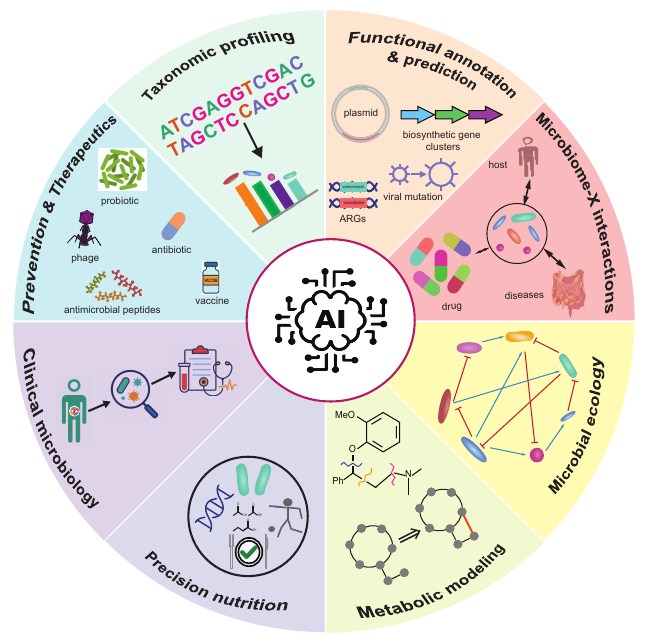}
  \caption{\textbf{Application scenarios of AI in microbiology and microbiome research.} The application scenarios of AI in microbiology and microbiome research are diverse, spanning (1) taxonomic profiling (e.g., metagenome assembly, binning, taxonomic classification, and nanopore sequencing basecalling), (2) functional annotation and prediction (including gene prediction, antibiotic resistance gene and plasmid identification, biosynthetic gene clusters prediction, 16S rRNA copy number prediction and mutation/evolution prediction), (3) microbe-host/disease/drug interactions, (4) microbial ecology (such as microbial interaction networks, composition and colonization outcome predictions and keystone species identification), (5) metabolic modeling, (6) precision nutrition (e.g., nutrition profile correction, metabolomic profile prediction and personalized diet recommendation), (7) clinical microbiology (such as microorganism detection, identification and quantification, antimicrobial susceptibility evaluation and disease diagnosis, classification, and clinical outcome prediction), and (8) applications in prevention and therapeutics (including peptides identification and generation, probiotic mining, antibiotic discovery, phage therapy and vaccine design).}
  \label{fig: application}
\end{figure*}

There are numerous applications of AI techniques in microbiology/microbiome research. We can briefly group those applications into the following scenarios: taxonomic profiling, functional annotation \& prediction, microbe-X interactions, microbial ecology, metabolic modeling, precision nutrition, clinical microbiology, prevention \& therapeutics (Fig.\ref{fig: application}). For each application scenario, there are many specific tasks. In the following, we will present each of the specific tasks and the representative AI methods.

\begin{landscape}
\begin{longtable}{p{5.4cm} p{5.6cm} p{5.6cm} p{5.6cm} >{\raggedleft\arraybackslash}p{0.0cm}}
\caption{Representative AI methods for microbiome research.}\\
\hline
\textbf{Name} & \textbf{Key AI architecture} & \textbf{Input data} & \textbf{Output data} & \textbf{Ref}\\
\hline
\endfirsthead
\hline
\textbf{Name} & \textbf{Key AI architecture} & \textbf{Input data} & \textbf{Output data} & \textbf{Ref}\\
\hline
\endhead
\hline
\endfoot

\multicolumn{5}{l}{\textbf{Taxonomic profiling}}\\
\hline
\textbf{Metagenome assembly}\\
DeepMAsEd & CNN & Contigs & Contig misassembly probability & \cite{mineeva2020deepmased} \\
ResMiCo & ResNet (CNN) & Contigs & Contig misassembly probability & \cite{mineeva2023resmico} \\
CheckM2 & Hybrid ML & MAG genome bins & MAG quality & \cite{chklovski2023checkm2} \\
DeepCheck & CNN & MAG genome bins & MAG quality & \cite{wei2024deepcheck} \\
\textbf{Metagenome binning}\\
VAMB & VAE & k-mer composition and abundance & Metagenomic bins & \cite{nissen_improved_2021} \\
CLMB & Contrastive learning + VAE & Contigs & Metagenomic bins & \cite{zhang2022clmb} \\
SemiBin & Siamese NN & Contigs & Metagenomic bins & \cite{pan_deep_2022} \\
GraphMB & VAE + GNN & Contigs & Metagenomic bins & \cite{lamurias_metagenomic_2022} \\
COMEBin & Contrastive multiview NN & Contigs & Metagenomic bins & \cite{wang2024effective} \\
\textbf{Taxonomic classification}\\
DeepMicrobes & BiLSTM & DNA sequence & Taxonomic classification & \cite{liang_deepmicrobes_2020} \\
BERTax & DNA BERT & DNA sequence & Taxonomic classification & \cite{mock_taxonomic_2022} \\
\textbf{Nanopore sequencing basecalling}\\
Chiron & CNN + RNN + CTC & Raw signal & DNA sequence & \cite{teng2018chiron} \\
SACall & CNN/ResNet + CTC & Raw signal & DNA sequence & \cite{huang2020sacall} \\
Mincall  & CNN/ResNet + CTC & Raw signal & DNA sequence & \cite{miculinic2019mincall} \\
Halcyon & Inception-CNN + LSTM + attention & Raw signal & DNA sequence & \cite{konishi2021halcyon} \\
URnano & U-Net (+RNN) & Raw signal & DNA sequence & \cite{zhang2020nanopore} \\

\hline
\multicolumn{5}{l}{\textbf{Functional annotation \& prediction}}\\
\hline
\textbf{Gene prediction}\\
Meta-MFDL & Deep stacked MLP & Metagenomic fragments & Metagenomic genes  & \cite{zhang2017gene} \\
CNN-MGP & CNN ensemble & Metagenomic fragments & Metagenomic genes & \cite{al2019cnn} \\
\textbf{Antibiotic resistance genes identification}\\
DeepARG & MLP & Metagenomic sequences & ARG categories & \cite{arango2018deeparg} \\
HMD-ARG & CNN & Metagenomic sequences & ARG categories & \cite{li2021hmd} \\
HyperVR & Ensemble DL & Metagenomic sequences & ARG categories & \cite{ji2023hypervr} \\
ARGNet & Autoencoder + CNN (hybrid) & Metagenomic sequences & ARG categories & \cite{pei2024argnet} \\
ARG-BERT & Protein BERT & Metagenomic sequences & ARG categories & \cite{yagimoto2024prediction} \\
\textbf{Identification of mobile genetic elements}\\
PlasFlow & MLP (plasmid ID) & Contigs & Plasmid sequences & \cite{krawczyk_plasflow_2018} \\
Deeplasmid & LSTM + FC & DNA sequences & Plasmid sequences &  \cite{andreopoulos_deeplasmid_2022} \\
plASgraph2 & GNN & Assembly graph & Plasmid contigs & \cite{sielemann2023plasgraph2} \\
PPR-Meta & Bi-path CNN & Metagenomic assemblies & Plasmid fragments & \cite{fang2019ppr} \\
\textbf{Biosynthetic gene clusters prediction}\\
DeepBGC & Pfam2vec + BiLSTM & Raw sequences & BGC classes & \cite{hannigan2019deep} \\
e-DeepBGC & BiLSTM + enriched embeddings & Raw sequences & BGC classes & \cite{liu2022deep} \\
BiGCARP & CNN masked LM & Raw sequences & BGC classes & \cite{rios2023deep} \\
BGC-Prophet & Transformer LM & Raw sequences & BGC classes & \cite{lai2025deciphering} \\
\textbf{16S rRNA copy number prediction}\\
ANNA16 & Stacked MLP + SVM & 16S gene sequences & 16S rRNA copy number & \cite{miao2024deep} \\
\textbf{Mutation/evolution prediction}\\
EVEscape & VAE + biophysics & Historical sequences & Mutation probability & \cite{thadani2023learning} \\

\hline
\multicolumn{5}{l}{\textbf{Microbe–X interactions}}\\
\hline
\textbf{Microbe-microbe interactions}\\
MicrobialCommunities & Random forest & Microbial interaction networks & Predicted interaction edges & \cite{dimucci2018machine} \\
\textbf{Microbe-host interactions}\\
PIPR & Recurrent CNN & Protein sequences & PPI & \cite{chen2019multifaceted} \\
DeepPPISP & TextCNN & Protein sequences & PPI & \cite{zeng2020protein} \\
DeepInterface & 3D CNN & Protein sequences & PPI & \cite{balci2019deepinterface} \\
MaSIF & CNN & Protein sequences & PPI & \cite{gainza2020deciphering} \\
PECAN & Graph CNN & Protein sequences & PPI & \cite{pittala2020learning} \\
\textbf{Microbe-disease associations}\\
NinimHMDA & GCN & MDA database & MDA & \cite{ma2020ninimhmda} \\
LGRSH & Node2vec & MDA database & MDA & \cite{lei2020predicting} \\
BPNNHMDA & MLP & MDA database & MDA & \cite{li2020identifying} \\
MICAH & Deep matrix factorization & MDA database & MDA & \cite{liu2023explainable} \\
\textbf{Microbe-drug associations}\\
GARFMDA & Graph attention network & Microbe-drug association & Microbe-drug association & \cite{kuang2024novel} \\
GCNATMDA & GCN + Graph attention network & Microbe-drug association & Microbe-drug association & \cite{wang2024microbe} \\
LCASPMDA & Graph attention network & Microbe-drug association & Microbe-drug association & \cite{yang2024lcaspmda} \\
MCHAN & GCN with attention & Microbe-drug association & Microbe-drug association & \cite{li2024mchan} \\
MDSVDNV & Node2vec & Microbe-drug association & Microbe-drug association & \cite{tan2024mdsvdnv} \\
NMGMDA & Graph attention network & Microbe-drug association & Microbe-drug association & \cite{liang2024nmgmda} \\
OGNNMDA & Graph attention network & Microbe-drug association & Microbe-drug association & \cite{Liu2024stnmda} \\
STNMDA & Structure-Aware Transformer & Microbe-drug association & Microbe-drug association & \cite{fan2024stnmda} \\


\hline
\multicolumn{5}{l}{\textbf{Microbial ecology}}\\
\hline
\textbf{Microbial composition prediction}\\
cNODE & Neural ODE & Species collection & Microbial composition &  \cite{michelmata_predicting_2022} \\
MicrobeGNN & GNN & Bacterial genome & Microbial composition & \cite{ruaud2024modelling} \\
\textbf{Keystone species identification}\\
DKI & Neural ODE & Microbial composition & Keystoneness & \cite{wang2024identifying} \\
\textbf{Microbial dynamics prediction}\\
MiRNN & Tailored RNN & Dynamic community data & Metabolite concentrations & \cite{thompson2023integrating} \\
MTSF-DG & GNN & Historical timeseries & Future timeseries & \cite{zhao2023multiple} \\

\textbf{Control of microbial communities}\\
ROCC & RL & Current microbial population levels & Optimal feed actions & \cite{treloar2020deep} \\

\textbf{Microbiome data simulation, augmentation, and imputation}\\
MB-GAN & GAN & Microbial abundances & Microbial abundances & \cite{rong_mb-gan_2021} \\
mbSparse & Autoencoder + CVAE & Raw taxonomic count matrix & Imputed taxonomic count matrix \cite{qi2025mbsparse} \\
\textbf{Microbial source tracking}\\
SourceTracker & Gibbs sampling & Taxonomic profile & Source contributions & \cite{knights2011bayesian} \\
FEAST & Fast expectation-maximization & Taxonomic profile & Source contributions & \cite{shenhav2019feast} \\
STENSL & Expectation-maximization & Taxonomic profile & Source contributions & \cite{an2022stensl} \\
ONN4MST & Ontology-aware Neural Network (ONN) & Taxonomic profile & Source contributions & \cite{zha_ontology-aware_2022} \\

\hline
\multicolumn{5}{l}{\textbf{Metabolic Modeling}}\\
\hline
\textbf{Gap filling: inferring missing reactions}\\
AGORA2 & GEM & Microbial genome sequences & Metabolic capabilities & \cite{heinken2023genome} \\
CHESHIRE & GCN & Metabolic networks & Candidate reactions & \cite{chen2023teasing} \\
\textbf{Retrosynthesis: breaking down a target molecule}\\
RetroPath RL & RL & Target compound, reaction-rule set & Predicted biosynthetic pathways \cite{koch2019reinforcement} \\
\textbf{Model microbial metabolism}\\
SPAM-DFBA & RL & Current community statet & Flux-regulation actions \cite{ghadermazi2024microbial} \\

\hline
\multicolumn{3}{l}{\textbf{Precision nutrition}}\\
\hline
\textbf{Nutrition profile correction}\\
METRIC & MLP & Gut microbiome, noised dietary profiles & Corrected dietary profiles & \cite{wang2024microbiome} \\
\textbf{Metabolomic profile prediction}\\
MiMeNet & MLP & Microbiome composition & Metabolome composition & \cite{reiman2021mimenet} \\
MelonnPan & linear regression & Microbiome composition & Metabolome composition & \cite{mallick2019predictive} \\
NED & MLP & Microbiome composition & Metabolome composition & \cite{le2020deep} \\
mNODE & neural ODE & Microbiome composition & Metabolome composition & \cite{wang2023predicting} \\
\textbf{Personalized diet recommendation}\\
McMLP & MLP & Baseline microbial compositions, dietary intervention strategy, baseline metabolomic profiles & Metabolomic profiles & \cite{wang2025predicting} \\

\hline
\multicolumn{5}{l}{\textbf{Clinical Microbiology}}\\
\hline
\textbf{Microorganism detection, identification and quantification}\\
BIOINTEL & CNN & Microscopy imaging & Bacteria type prediction & \cite{ramesh2024biointel} \\
\textbf{Antimicrobial susceptibility evaluation}\\
ANN & MLP & Mutation patterns & Resistance to the protease inhibitor lopinavir & \cite{wang2003enhanced} \\
GoPhAST-R & RF & NanoString data  & Probability of resistance to antibiotics & \cite{bhattacharyya2019simultaneous} \\
CART & Classification and Regression Trees & Pyrazinamide MIC  & MIC threshold & \cite{gumbo2014pyrazinamide} \\
\textbf{Disease diagnosis, classification, and clinical outcome prediction}\\
Ph-CNN & Phylogeny-based CNN & Taxonomic profile & Sample categories & ~\cite{fioravanti_phylogenetic_2018} \\
PopPhy-CNN & Phylogeny-based + abundance + CNN & Taxonomic profile & Sample categories & \cite{reiman2020popphy} \\
taxoNN & Phylogeny cluster-based CNN & Taxonomic profile & Sample categories & \cite{sharma2020taxonn} \\
MDeep & Phylogeny-based CNN & Taxonomic profile & Sample categories & \cite{wang_novel_2021} \\
PM-CNN & Phylogeny-based CNN & Taxonomic profile & Sample categories & ~\cite{wang2024pm} \\
DeepMicro & Autoencoders + MLP & Taxonomic profile & Sample categories & \cite{liang_deepmicrobes_2020} \\
GDmicro & GCN & Taxonomic profile & Sample categories & \cite{liao2023gdmicro} \\
MLM & Transformer & Taxonomic profile & Sample categories & \cite{pope2023learning} \\
MSFT-transformer & Multistage fusion tabular Transformer & Taxonomic and functional profiles & Sample categories & \cite{wang2025msft} \\
\textbf{Integration of various feature types}\\
MDL4Microbiome & MLP & Taxonomic and functional profiles & Sample categories & \cite{lee_multimodal_2022} \\

\hline
\multicolumn{5}{l}{\textbf{Prevention and Therapeutics}}\\
\hline
\textbf{Vaccine design}\\
Vaxi-ML & XGBoost & Protein features & Protective-antigen prediction & \cite{ong2020vaxign} \\
Vaxi-DL & MLP & Protein sequences & Protective-antigen prediction & \cite{rawal2022vaxi} \\
\textbf{Probiotic mining}\\
iProbiotics & SVM & Whole-genome sequences & Probiotic property prediction per genome \cite{sun_iprobiotics_2022} \\
metaProbiotics & Random forest & Metagenomic bins & Per-bin probiotic score \cite{wu2024metaprobiotics} \\
\textbf{Antibiotic discovery}\\
MPNN & GCN & Molecular graphs & Predicted molecular properties & \cite{gilmer2017neural} \\
Chemprop & DL GCN + MLP & Molecular graphs & Predicted molecular properties & \cite{heid2023chemprop} \\
\textbf{Antimicrobial peptides identification \& design}\\
MLBP & CNN & Peptide sequence & Peptide categories & \cite{tang2022identifying} \\
AMPGANv2 & cGAN & Peptide sequence & Potent AMPs & \cite{van2021ampgan} \\
EvoGradient & CNN, LSTM, Transformer & Peptide sequence & Potent AMPs & \cite{wang2025explainable} \\
deepAMP & CVAE & Peptide sequence & Potent AMPs & \cite{li2024foundation} \\
\textbf{Phage therapy}\\
INHERIT & DNA BERT & DNA contigs & Phage vs. non-phage classification & \cite{bai2022identification} \\
PHACTS & RF & Proteome profiles of bacteriophages & Phage lifestyle classification & \cite{mcnair2012phacts} \\
DeePhage & CNN & DNA fragments & Phage lifestyle classification & \cite{wu_deephage_2021} \\
DeePhafier  & Self-attention & Phage protein features & Phage lifestyle classification & \cite{miao2024deephafier} \\
PhaTYP & Transfomer & Phage genome & Phage lifestyle classification & \cite{shang2023phatyp} \\
PHIAF & GAN & Phage and host genome & Predicted phage-host interaction & \cite{li2022phiaf} \\
PHPGAT & GAT & Integrated phage-phage, host-host and phage-host links & Predicted phage-host interaction & \cite{liu2025phpgat} \\
SpikeHunter & ESM-2 & Metagenomic contigs & Phage categories & \cite{yang2024large} \\
PhANNs & MLP & Phage protein sequences & Protein structural class & \cite{cantu2020phanns} \\
VirionFinder & CNN & Viral protein sequences & Virion vs. non-virion classification & \cite{fang2021virionfinder} \\
DeePVP & CNN & Phage protein sequences & Virion vs. non-virion classification & \cite{fang2022deepvp} \\
PhaVIP & Transformer & Phage protein sequences & Virion vs. non-virion classification & \cite{shang2023phavip} \\
ESM-PVP & ESM-2 & Phage protein sequences & Virion vs. non-virion classification & \cite{li2023esm} \\
DeepLysin & Ensemble model & Phage protein sequences & Antibacterial activity score per lysin & \cite{zhang2024discovery} \\
DeepMineLys & CNN & Lysins protein sequences & Predicted phage lysins & \cite{fu2024deepminelys} \\
\end{longtable}
\end{landscape}

\subsection{Taxonomic Profiling}

A fundamental goal of microbiology and microbiome research is determining the compositions of microbial communities, i.e., identifying and quantifying different types of microorganisms (such as bacteria, fungi, viruses, and archaea) present in a given sample. This involves analyzing their relative abundances and diversity, often using DNA sequencing techniques. Currently, three generations of DNA sequencing techniques are available for microbiome research. The first-generation sequencing utilizes the chain termination method, offering read lengths of 500-1000 base pairs~\cite{shendure2017dna}. Second-generation sequencing, also known as next-generation sequencing (NGS), includes methods such as pyrosequencing, sequencing by synthesis, and sequencing by ligation, with read lengths ranging between 50 and 500 bp~\cite{knight2018best,pinto2024sequencing,quince2017shotgun}. Two key NGS applications in microbiome research are (1) amplicon sequencing, which targets small fragments of one or two hypervariable regions of the 16S rRNA gene (for archaea and bacteria) or 18S rRNA gene (for fungi); and (2) metagenomic shotgun sequencing, which comprehensively samples all genes in all organisms present in a given community. 
The third-generation sequencing performs single-molecule sequencing, offering long reads with lengths reaching tens of kilobases on average~\cite{schadt2010window}. In the following, we discuss applications of AI techniques in various aspects of taxonomic profiling. 

\phantomsection
\subsubsection{Metagenome assembly}
Metagenomics refers to the direct study of the entire genomic information contained in a microbial community. Metagenomics avoids isolating and culturing individual microorganisms in a community and provides a way to study microorganisms that cannot be isolated and cultured.  There are two main approaches for processing metagenomic sequencing data: (1) assembly-based and (2) reference database-based. The goal of the assembly-based approach is to construct and annotate the so-called metagenome-assembled genomes (MAGs)~\cite{setubal2021metagenome}. The construction of MAGs includes two main steps: assembly and binning. Assembly refers to the process of reconstructing longer sequences (contigs) from DNA sequencing reads, which can vary in length and type depending on the technology used. This involves piecing together overlapping reads—using short-read, linked-read, long-read, or hybrid assemblers—to form contiguous sequences that represent parts of the genomes present in the microbial community~\cite{zhang2023benchmarking,zhang2024exploring}.  

Machine learning has been widely used in the quality control of metagenomic assembly. Many factors (e.g., sequencing errors, variable coverage, repetitive genomic regions, etc.) can produce misassemblies. For taxonomically novel genomic data, detecting misassemblies is very challenging due to the lack of closely related reference genomes. Machine learning methods can identify misassembled contigs in a reference-free manner. Representative methods include DeepMAsEd~\cite{mineeva2020deepmased}, ResMiCo~\cite{mineeva2023resmico} CheckM2~\cite{chklovski2023checkm2}, and DeepCheck~\cite{wei2024deepcheck}. DeepMAsEd is based on CNN. Denote a contig as a sequence of nucleotides. At each position in the sequence, the concatenation of two types of information (raw sequence and read-count features) yields the input vector. To train and evaluate DeepMAsEd, one can generate a synthetic dataset of contigs, read counts, and binary assembly quality labels. As an extension of DeepMAsEd, ResMiCo is based on ResNet, a variant of CNN. The key feature of ResNet is the introduction of skip connections, which effectively solves the degradation problem of deep neural networks~\cite{he_deep_2016}. Compared to DeepMAsEd, ResMiCo leveraged a much more informative input vector computed from raw reads and contigs. CheckM2 is a machine learning framework to predict genome quality of MAGs~\cite{chklovski2023checkm2}. CheckM2 uses both CNN and the gradient boosted (GB) decision tree for completeness prediction, and uses the GB model to predict contamination. Since CheckM2 employs two separate models to predict the completeness and contamination of MAGs, it may overlook the inherent correlation between these two tasks. 
DeepCheck is a multitasking deep learning framework based on ResNet and the attention mechanism to simultaneously predict MAG completeness and contamination~\cite{wei2024deepcheck}. 
We notice that those methods typically use a carefully designed input vector. 
For example, ResMiCO transforms raw reads and contigs into carefully selected positional features, such as coverage (number of reads aligned to position),  number of single-nucleotide variants (SNVs), mean alignment score, etc. 
CheckM2 and DeepCheck calculate feature vectors from the genome properties of synthetic MAGs, e.g., the genome length, number of coding sequences and individual amino acid counts, as well as annotation of predicted proteins using KEGG (the Kyoto Encyclopedia of Genes and Genomes).
It would be interesting to explore if we can use a more advanced deep learning architecture (e.g., the Transformer) or a hybrid learning approach (e.g., CNN + RNN) to directly deal with the raw
sequence data, avoiding the manual design of the input vector.

\phantomsection
\subsubsection{Metagenome binning}
Metagenomic binning involves grouping those assembled sequences into clusters (bins or MAGs) that correspond to different species or genomes~\cite{herazo2025review}. Metagenomic binning helps us identify and categorize different microorganisms present in a metagenomic sample, even if they are not fully assembled into complete genomes. There are many methods for metagenomic binning~\cite{sedlar2017bioinformatics,yang2021review,lettich2024genomeface,lamurias2023graph}. Several binning methods are based on deep learning, e.g., VAMB~\cite{nissen_improved_2021}, CLMB~\cite{zhang2022clmb}, SemiBin~\cite{pan_deep_2022}, GraphMB~\cite{lamurias_metagenomic_2022}, and COMEBin~\cite{wang2024effective}. VAMB (Variational Autoencoders for Metagenomic Binning) uses VAE to encode sequence coabundance and $k$-mer distribution information, and clusters the resulting latent representation into genome clusters and sample-specific bins~\cite{nissen_improved_2021}. As an extension of VAMB, CLMB (Contrastive Learning framework for Metagenome Binning) can efficiently eliminate the disturbance of noise and produce more stable and robust results~\cite{zhang2022clmb}. CLMB is based on contrastive learning, an machine learning approach that focuses on extracting meaningful representations by contrasting positive and negative instances~\cite{zhang2022clmb}. SemiBin employs deep siamese neural networks to exploit the information in reference genomes, while retaining the capability of reconstructing high-quality bins that are outside the reference dataset~\cite{pan_deep_2022}. Here, a siamese neural network (a.k.a. twin neural network) is a neural network that uses the same weights while working in tandem on two different input vectors to compute comparable output vectors~\cite{chicco2021siamese}. GraphMB integrates GCN with assembly graphs to improve binning accuracy~\cite{lamurias_metagenomic_2022}. It models each contig using VAE for feature generation and aggregates these features using a GCN. This method accounts for read coverage in its loss function and uses iterative medoid clustering to finalize the binning. COMEBin is the latest metagenomic binning method~\cite{wang2024effective}. This method is based on contrastive multiview representation learning. It introduces a data augmentation approach that generates multiple views for each contig, enabling contrastive learning and yielding high-quality representations of the heterogeneous features. Moreover, it incorporates a ``Coverage module'' to obtain fixed-dimensional coverage embeddings, which enhances its performance across datasets with varying numbers of sequencing samples. It also adapts an advanced community detection algorithm, Leiden, specifically for the binning task, considering single-copy gene information and contig length. COMEBin outperformed VAME and SemiBin on various simulated and real datasets, especially in recovering near-complete genomes from real environmental samples.

\phantomsection
\subsubsection{Taxonomic classification} 
All the methods discussed in the previous section are assembly-based metagenomic analysis methods. There are also many metagenomic analysis methods based on reference databases. In particular, those methods used for microbial classification and abundance estimation are also known as metagenomic profilers, which can be grouped into three categories based on the type of reference data~\cite{simon2019benchmarking}: (1) DNA-to-DNA methods (such as Bracken~\cite{lu2017bracken}, Kraken~\cite{wood2014kraken,wood2019improved}, and PathSeq~\cite{kostic2011pathseq}), which compare sequence reads with comprehensive genomes; (2) DNA-to-Protein methods (such as Diamond~\cite{buchfink2015fast}, Kaiju~\cite{menzel2016fast}, and MMSeqs~\cite{hauser2016mmseqs,steinegger2017mmseqs2}), which compare sequence reads with protein-coding DNA; (3) DNA-to-Marker methods (such as MetaPhlAn~\cite{segata2012metagenomic,truong2015metaphlan2,beghini2021integrating,blanco2023extending} and mOTUs~\cite{sunagawa2013metagenomic,milanese2019microbial}), whose reference databases only contain specific gene families. It has been pointed out that the output of the first two categories is the sequencing abundance of species (without correction for genome size and copy number), while the output of the third category is the species abundance in a taxonomic or ecological sense~\cite{sun2021challenges}. Given these different types of relative abundances, benchmarking metagenomic profilers remains a big challenge~\cite{sun2021challenges}.

These metagenomic profilers query DNA sequences in reference databases based on the concept of homology, which refers to the similarity between sequences of DNA, RNA, or protein that is due to shared ancestry. Obviously, those methods are largely affected by the quality of the reference database. A rather optimistic estimate suggests that the number of reference genomes in current comprehensive databases (such as RefSeq) may account for less than 5.319$\%$ of all species~\cite{louca2019census}. This explains why homology-based methods sometimes work poorly.

Deep learning techniques provide an alternative solution. These deep learning methods do not rely on similar sequences to exist in the reference database, and they allow for the modeling of complex correspondences between DNA sequences and corresponding species classifications. In these deep learning methods, DNA sequences are usually encoded into numeric matrices first, e.g., converting a sequence into a one-hot matrix or embedding the $k$-mers into a representative matrix. For example, DeepMicrobes is a deep learning method for taxonomic classification of short metagenomic sequencing reads~\cite{liang_deepmicrobes_2020}. In DeepMicrobes, DNA sequences are segmented into substrings, each mapped to a 100-dimensional embedding vector. These vectors are processed by a bidirectional LSTM and a self-attention layer, which prioritizes relevant $k$-mers (with $k=12$) for the classification task. The LSTM outputs are combined with attention scores to produce an output matrix that feeds into a classifier for final species and genus identification. DeepMicrobes outperforms traditional tools like Kraken~\cite{wood2014kraken}, Kraken2~\cite{wood2019improved} (where sequences are classified using the taxonomic tree), CLARK (using target-specific $k$-mer for classification)~\cite{ounit2015clark} in accuracy, but requires extensive computational resources and dataset sizes. Moreover, adding new species also necessitates retraining the entire network.

BERTax is another deep learning method for taxonomic classification. It classifies DNA sequences into three different classification levels, namely superkingdom (archaea, bacteria, eukaryotes, and viruses), phylum, and genus~\cite{mock_taxonomic_2022}. The novelty of BERTax is to assume DNA is a ``language'' and to classify the taxonomic origin based on this language understanding rather than by local similarity to known genomes in any database (i.e., homology). As its name suggests, BERTax is based on the state-of-the-art NLP architecture BERT (bidirectional encoder representations from transformers), which relies on a transformer employing the mechanism of self-attention. The training process of BERTax consists of two steps. First, BERT is pre-trained in an unsupervised manner, with the goal of learning the general structure of the genomic DNA ``language''. Second, the pre-trained BERT model is combined with a classification layer and fine-tuned for the specific task of predicting classification categories. It has been shown that BERTax is at least comparable to state-of-the-art methods when similar species are part of the training data. However, for the classification of new species, BERTax significantly outperforms any existing method. BERTax can also be combined with database approaches to further increase the prediction quality in almost all cases.

\phantomsection
\subsubsection{Nanopore sequencing basecalling}
Since its introduction, nanopore sequencing has revolutionized microbial community studies by enabling portable, real-time long-read sequencing (with reads longer than a few thousand bases) at low cost~\cite{ branton2008potential,ciuffreda2021nanopore}. Its long reads overcome key challenges in reconstructing complex genomic structures and assembling high-quality MAGs. Coupled with rapid sample-to-answer workflows and streamlined analysis pipelines, nanopore technology now offers an accessible and powerful platform for real-time, in-field monitoring of environmental microbiomes. The basic principle of nanopore sequencing is to pass an ionic current through a nanopore and measure the change in current when a biomolecule passes through or approaches the nanopore. Information about the change in current can be used to identify the molecule, a process often referred to as basecalling. There are two challenges in basecalling. First, the current signal level is most dominantly influenced by the several nucleotides that reside inside the pore at any given time, rather than a single base. Second, DNA molecules do not translocate at a constant speed. Basecalling is conceptually similar to speech recognition. Both processes involve interpreting complex signals to extract meaningful sequences-DNA bases in the case of basecalling, and spoken words in the case of speech recognition. Much like the evolution of speech recognition methods, computational methods for basecalling have evolved from statistical tests to hidden Markov models and finally deep learning models. Those methods are often referred to as basecallers. 

Various deep learning models have been developed for basecalling. Chiron is the first deep learning model that can translate raw electrical signal directly to nucleotide sequence~\cite{teng2018chiron}. It applied a CNN to extract features from the raw signal, an RNN to relate such features in a temporal manner, and a connectionist temporal classification (CTC) decoder to create the nucleotide sequence. Here, CTC enabled us to generate a variant length base sequence for a fixed-length signal window through output-space searching, avoiding explicit segmentation for basecalling from raw signals. Similar to the Chiron architecture, SACall~\cite{huang2020sacall} integrated CNN with Transformer (Lite Transformer or LSTM) and CTC. Similar approaches are also implemented in CATCaller~\cite{lv2020end} and Bonito~\cite{xu2021fast}. Mincall~\cite{miculinic2019mincall} (or Causalcall~\cite{zeng2020causalcall}) directly integrated ResNet (or causal dilated CNN) with CTC. Halcyon used a different architecture ~\cite{konishi2021halcyon}. It combines a novel inception-block-based CNN module, an LSTM-based encoder, and an LSTM-based decoder using an attention mechanism. The inception-block-based CNN module aims to extract local features of input raw signal and reduce the dimension of the input timestep axis. The LSTM-based encoder captures long-time dependencies in the timestep dimension and deals with the variable lengths of inputs. The attention mechanism allows the decoder to focus on specific parts of the input sequence when generating each element of the output sequence.

All those methods mentioned so far treat basecalling as a sequence labeling task.  URnano formalized the basecalling as a multi-label segmentation task that splits raw signals and assigns corresponding labels~\cite{zhang2020nanopore}. In particular, URnano used a U-Net with integrated RNNs. Here, U-Net is a u-shaped CNN architecture that was originally designed for biomedical image segmentation~\cite{DBLP:journals/corr/RonnebergerFB15}. 

Benchmarking and architecture analysis of these deep learning-based basecallers show that: (1) the conditional random field (CRF) decoder is vastly superior to CTC; (2) complex convolutions are most robust, but simple convolutions are still very competitive; (3) LSTM is superior to Transformer and is depth dependent~\cite{pages2023comprehensive}. The reason why the attention mechanism in Transformer is not beneficial for basecalling could be the temporal relationships in the electric signal are local enough so that LSTM is sufficient for the task.

\subsection{Functional Annotation \& Prediction}
\phantomsection
\subsubsection{Gene prediction}
After carefully selecting MAGs from the metagenome assembly, we need to identify and annotate genes by recognizing potential coding sequences within MAGs~\cite{yang2021review}. This can be achieved by two types of methods: model-based methods (e.g., MetaGeneMark~\cite{zhu2010ab}, Glimmer-MG~\cite{kelley2012gene} and FragGeneScan~\cite{rho2010fraggenescan} using hidden Markov models, and Prodigal~\cite{hyatt2010prodigal}, MetaGene~\cite{noguchi2006metagene}, MetaGeneAnnotator~\cite{noguchi2008metageneannotator} using dynamic programming); and deep learning-based methods (e.g., Meta-MFDL~\cite{zhang2017gene}, CNN-MGP~\cite{al2019cnn}, and Balrog~\cite{sommer2021balrog}). Meta-MFDL generates a representation vector by integrating various features (e.g., single codon usage, mono-amino acid usage, etc.), and subsequently trains a deep stacking network to classify coding and non-coding ORFs. Here, the deep stacking network is composed of a series of modules with the same or similar structure stacked together. For Meta-MFDL, the authors used a simple MLP with only one hidden layer for each module. The ``stacking'' is completed by combining the outputs of all previous modules with the original input vector to form a new ``input'' vector as the input of the next module. CNN-MGP utilizes CNNs to automatically learn features of coding and non-coding ORFs from the training dataset and predict the probability of ORFs in MAGs. The authors extracted ORFs from each metagenomics fragment and encoded ORFs numerically. Then they built 10 CNN models for classification. Finally, they used 10 CNN classifiers to approximate the gene probability for the candidate ORFs, and used a greedy algorithm to select the final gene set. Balrog uses a TCN to predict genes based on a large number of diverse microbial genomes. The authors used the state of the last node of the linear output layer of the TCN as representative of the binary classifier, with a value close to 1 predicting a protein-coding gene sequence and 0 predicting an out-of-frame sequence. It is not clear which of those gene prediction methods is the best. Systematic benchmarking is necessary.

\phantomsection
\subsubsection{Antibiotic resistance genes identification}
Antibiotics become less effective as bacterial pathogens develop and spread resistance over time. This has led to the antibiotic resistance crisis, e.g., resistance may involve most or even all the available antimicrobial options~\cite{rossolini2014update}. It has been estimated that antibiotic resistance could cause over 10 million deaths annually by 2050 if no significant action is taken. The economic costs associated with these outcomes could also reach approximately 100 trillion USD globally~\cite{de2016will}. Some particular ecosystems, for instance, wastewater, have been considered reservoirs and environmental suppliers of antibiotic resistance due to the spreading of antibiotic resistance gene transfer between different bacterial species~\cite{karkman2018antibiotic,zhang2009antibiotic}. Computational methods that can help identify potential resources of novel antibiotic resistance genes (ARGs) are particularly crucial.

DeepARG is a deep learning approach for predicting ARGs from metagenomic data~\cite{arango2018deeparg}.
First, genes in Uniprot were aligned to the CARD and ARDB databases using DIAMOND to obtain the dissimilarity representation, e.g., bit score after normalization so that scores close to 0 represent small distance or high similarity, and scores around 1 represent distant alignments. The final feature matrix indicates the sequence similarity of the Uniprot genes to the ARDB and CARD genes. The feature matrix was fed into four dense fully connected hidden layers and a SoftMax output layer to predict the probability of the input sequence against each ARG category.
HMD-ARG is an end-to-end hierarchical multi-task deep learning framework for ARG annotation~\cite{li2021hmd}. 
HMD-ARG used a CNN model where each sequence composed of 23 characters representing different amino acids was converted into one-hot encoding. Those sequence encodings were fed into six convolutional layers and four pooling layers to detect important motifs and aggregate local and global information across input sequences. The outputs of the last pooling layer were flatted and fed into three fully connected layers and a Softmax layer to predict final labeling~\cite{li2021hmd}. 
HyperVR is a hybrid deep ensemble learning method that can simultaneously predict virulence factors and ARGs~\cite{ji2023hypervr}. 
ARGNet is a two-stage deep learning approach that incorporates an unsupervised deep learning model autoencoder to first identify ARGs from the input genomic sequences and then uses a supervised deep learning model CNN to predict the antibiotic resistance category for sequences determined as ARGs by the autoencoder~\cite{pei2024argnet}. This hybrid learning approach enables a more efficient discovery of both known and novel ARGs. It was shown that ARGNet outperformed DeepARG and HMD-ARG in most of the applications and reduced inference runtime by up to 57$\%$ relative to DeepARG.

Ground-breaking LLMs initially created for NLP have found success in predicting protein functions. These models, referred to as protein language models (PLMs), excel at generating intricate semantic representations that forge meaningful links between gene sequences and protein functions~\cite{rives2021biological,elnaggar2021prottrans,brandes2022proteinbert}. FunGeneTyper is a PLM-based deep learning framework designed for accurate and scalable prediction of protein-coding gene functions~\cite{zhang2024highly}. This framework includes two interconnected deep learning models: FunTrans and FunRep. While these models share a similar architecture, they are tailored for classifying functional genes at type and subtype levels, respectively. Both models utilize modular adapter-based architectures, incorporating a few additional parameters for efficient fine-tuning of extensive PLMs. Specifically, utilizing the ESM-1b model (a large-scale PLM built on a 33-layer transformer architecture~\cite{rives2021biological}), adapters are inserted into each transformer layer, serving as individual modular units that introduce new weights tuned for specific tasks. FunGeneTyper has shown exceptional performance in classifying ARGs and virulence factor genes. More significantly, it is a flexible deep learning framework that can accurately classify general protein-coding gene functions and aid in discovering numerous valuable enzymes. ARG-BERT is another PLM-based model for predicting ARG resistance mechanisms~\cite{yagimoto2024prediction}. Leveraging ProteinBERT~\cite{brandes2022proteinbert} (a BERT-based PLM), ARG-BERT includes an input layer for ARG sequences, and an output layer for predicting six resistance mechanism labels: antibiotic efflux, antibiotic inactivation, antibiotic target alteration, antibiotic target protection, antibiotic target replacement, and others. ARG-BERT was fine-tuned using the HMD-ARG database~\cite{li2021hmd} (composed of 17,282 high-quality sequences, coupled with labels of 15 antibiotic classes, 6 underlying resistance mechanisms, and their mobility), as well as a custom-built low-homology dataset. Subsequent attention analysis performed on the fine-tuned model (with HMD-ARG database)  suggests that ARG-BERT takes into account biologically relevant features (e.g., antibiotic-binding sites and evolutionarily conserved regions) in making predictions.

\phantomsection
\subsubsection{Identification of mobile genetic elements}
Mobile genetic elements are a type of genetic material that can move around within a genome or be transferred from one species to another. They are often referred to as selfish genetic elements, because they have the ability to promote their own transmission at the expense of other genes in the genome. They are found in all organisms. The set of mobile genetic elements in an organism is called a mobilome, including plasmids, viruses, transposons, integrons, introns, etc.

Plasmids are small, typically circular DNA molecules that are found in many microorganisms. They play an important role in microbial ecology and evolution through horizontal gene transfer, antibiotic resistance, and ecological interaction, etc. Identifying plasmid sequences from microbiome studies can provide a unique opportunity to study the mechanisms of plasmid persistence, transmission, and host specificity~\cite{andreopoulos_deeplasmid_2022}. Many classical machine learning methods have been proposed for plasmid identification, e.g., cBar~\cite{zhou2010cbar} based on sequential minimal optimization, PlasClass~\cite{pellow2020plasclass} using Logistic Regression, PlasmidVerify~\cite{antipov2019plasmid} using Naïve Bayesian classifier, PlasForest~\cite{pradier2021plasforest}, Plasmer~\cite{zhu2023plasmer}, Plasmidhunter~\cite{tian2024plasmidhunter}, RFPlasmid~\cite{van2021rfplasmid} and SourceFinder~\cite{aytan2022sourcefinder} using Random Forest. Several deep learning methods have also been developed for plasmid identification. For example, PlasFlow employs MLP for the identification of bacterial plasmid sequences in environmental samples~\cite{krawczyk_plasflow_2018}. It can recover plasmid sequences from assembled metagenomes without any prior knowledge of the taxonomical or functional composition of samples with high accuracy. Deeplasmid is another deep learning method for distinguishing plasmids from bacterial chromosomes based on the DNA sequence~\cite{andreopoulos_deeplasmid_2022}. It leverages both LSTM and fully connected layers to generate features, which are then concatenated and passed to another block of fully connected layers to generate the final output--the Deeplasmid score. The higher the score is for the sequence, the more likely it is to be a true plasmid. plASgraph2 is a new deep learning method for identifying plasmid contigs in fragmented genome assemblies built from short-read data~\cite{sielemann2023plasgraph2}. The innovation of plASgraph2 lies in its use of GCN and the assembly graph to propagate information from neighboring nodes, resulting in more accurate classification. The GCN model consists of a set of graph convolutional layers designed to propagate information from neighboring contigs within the assembly graph. plASgraph2 generates two scores for each graph node: a plasmid score and a chromosomal score, which are used to assess whether a contig is likely derived from a plasmid, chromosome, or both. 


Recently, deep learning methods have been developed to simultaneously identify both viruses and plasmids, the two major components of the mobilome. For example, PPR-Meta is the first tool that can simultaneously identify phage and plasmid fragments from metagenomic assemblies efficiently and reliably~\cite{fang2019ppr}. PPR-Meta leveraged a novel architecture, Bi-path CNN, to improve the performance for short fragments. The Bi-path CNN leverages both base and codon information to enhance performance: the ``base path'' is effective for extracting sequence features of noncoding regions, while the ``codon path'' is useful for capturing features of coding regions. 

\phantomsection
\subsubsection{Biosynthetic gene clusters prediction}


Natural products are chemical compounds that serve as the foundation for numerous therapeutics in the pharmaceutical industry~\cite{dias2012historical}. In microbes, microbial secondary metabolites (an important source of natural products) are produced by clusters of colocalized genes known as biosynthetic gene clusters (BGCs)~\cite{wang2019gut}. Advances in high-throughput sequencing have led to a surge in the availability of complete microbial isolate genomes and metagenomes, offering a great opportunity to discover a vast number of new BGCs. BGC identification primarily employs rule-based or machine learning approaches. Rule-based methods (e.g., antiSMASH~\cite{blin2023antismash} and PRISM~\cite{skinnider2017prism}) excel at detecting known BGC categories (e.g., alkaloids, nonribosomal peptides, polyketides, ribosomally synthesized and post-translationally modified peptides (RiPP), saccharides, terpenes) but struggle with novel BGCs. Many machine learning methods~\cite{cimermancic2014insights,de2019neuripp,merwin2020deepripp,hannigan2019deep,liu2022deep,rios2023deep,lai2025deciphering} have been developed to overcome the limitations of rule-based methods. In particular, deep learning techniques have been very useful in this endeavor.

For example, DeepBGC and its extension employ (1) Pfam2vec (a word2vec-like word embedding model, which is a shallow neural network with a single hidden layer);
(2) a Bidirectional LSTM (a classical RNN), which offers the advantage of capturing short- and long-term dependencies between adjacent and distant genes \cite{hannigan2019deep}. e-DeepBGC still leverages those neural networks, but improves DeepBGC in the following aspects \cite{liu2022deep}. First, e-DeepBGC employs Pfam names, Pfam domain summary, Pfam domain clan information. This additional information is used to create new embedding of each Pfam domain by providing more biological information than that encoded by Pfam2vec which only uses the Pfam names. Second, a novel data augmentation step is introduced to overcome the limited number of BGCs with known functional classes.

BiGCARP is a self-supervised neural network masked language model~\cite{rios2023deep}. It is based on the convolutional autoencoding representations of proteins (CARP), a masked language model of proteins. That’s why it is called Biosynthetic Gene CARP (or BiGCARP). The CARP is based on  CNN, and has been shown to be competitive with transformer-based models for protein sequence pretraining~\cite{yang2024convolutions}.

BGC-Prophet is a transformer-based language model for BGC prediction and classification~\cite{lai2025deciphering}. It employs ESM2 for sequence-specific gene embeddings and a transformer encoder with classifiers for BGC detection and product classification. This approach enhances accuracy and efficiency in identifying both known and novel BGCs by capturing location-dependent gene relationships. In particular, BGC-Prophet is several orders of magnitude faster than existing methods, e.g., DeepBGC, enabling pan-phylogenetic screening and whole-metagenome screening of BGCs.


\phantomsection
\subsubsection{16S rRNA copy number prediction}
The 16S rRNA gene is highly conserved across different bacterial species but contains hypervariable regions that provide species-specific signatures. By sequencing these regions, we can determine the composition and diversity of bacterial communities in various environments. Yet, different bacterial species can have varying numbers of 16S rRNA gene copies (ranging from 1 to 21 copies/genome), which can lead to biases in quantifying microbial communities if not accounted for~\cite{klappenbach2000rrna}. To accurately estimate the relative abundance of bacterial species in a microbiome sample, we need to adjust the proportion of 16S rRNA gene read counts by the inverse of the 16S rRNA gene copy number. Experimentally measuring the 16S rRNA gene copy numbers through whole genome sequencing or competitive PCR is expensive and/or culture-dependent. To resolve this limitation, based on the hypothesis that 16S rRNA gene copy number correlates with the phylogenetic proximity of species, many bioinformatics tools have been developed to infer 16S rRNA gene copy numbers from taxonomy or phylogeny~\cite{kembel2012incorporating,angly2014copyrighter,stoddard2015rrn,douglas2020picrust2}. Yet, an independent assessment demonstrated that regardless of the method tested, 16S rRNA gene copy numbers could only be accurately predicted for a limited fraction of taxa~\cite{louca2018correcting}.
   
Recently, a deep learning-based method ANNA16 was developed to predict 16S rRNA gene copy numbers directly from DNA sequences, avoiding information loss in taxonomy classification and phylogeny~\cite{miao2024deep}.  Essentially, ANNA16 treats the 16S GCN prediction problem as a regression problem. A stacked ensemble model (mainly consisting of MLP and SVM) is the core of ANNA16. The 16S rRNA gene sequences were first preprocessed with K-merization. The resulting k-mer counts (with k=6) and the existing 16S rRNA gene copy number data (retrieved from rrnDB database) were used to train the stacked ensemble model. Based on 27,579 16S rRNA gene sequences and copy number data, it has been shown that ANNA16 outperforms previous methods (i.e., rrnDB, CopyRighter, PAPRICA, and PICRUST2). We expect that in the near future more deep learning-based methods will be developed to solve this fundamental problem in microbiology and microbiome research.

\phantomsection
\subsubsection{Mutation/evolution prediction}
Predicting evolution has been a longstanding objective in evolutionary biology, offering significant implications for strategic pathogen management, genome engineering, and synthetic biology. In microbiology, evolution prediction has been studied for several microorganisms. For instance, Wang et al. used the evolutionary histories of \textit{Escherichia coli} to train an ensemble predictor to predict which genes are likely to have mutations given a novel environment~\cite{wang2018predicting}. To achieve that, they first created a training dataset consisting of more than 15,000 mutation events for \textit{E. coli} under 178 distinct environmental settings reported in 95 publications.  For each mutation event, they recorded its genome position with respect to a reference genome and the mutation event type (e.g., single-nucleotide polymorphisms (SNPs), deletions, insertions, amplifications, inversions). Then, they integrated a deep learning model MLP and two classical machine learning models, Support Vector Machine and Naive Bayes, to build an ensemble predictor to predict the mutation probability of any given gene under a new environment. The input of the ensemble predictor consists of 83 binary variables (features) that capture attributes related to the strain, medium, and stress from experiments. The model output is a binary variable that captures the presence/absence of mutation(s) in any given gene, computed from the predicted probability of this gene’s mutation event. This work clearly illustrated how the evolutionary histories of microbes can be utilized to develop predictive models of evolution at the gene level, clarifying the impact of evolutionary mechanisms in specific environments. One limitation of this approach is that those 83 features were manually selected, which relies on domain knowledge. 

Another interesting work is EVEscape, a generalizable modular framework that can predict viral mutations based on pre-pandemic data~\cite{thadani2023learning}. It has been shown that EVEscape, if trained on sequences available before 2020, is as accurate as high-throughput experimental scans in predicting pandemic variation for SARS-CoV-2 and is generalizable to other viruses (such as influenza, HIV, Lassa, and Nipah). The EVEscape framework is based on the assumption that the probability that a viral mutation will induce immune escape is the joint probability of three independent events: (1) this mutation will maintain viral fitness (‘fitness’ term); (2) the mutation will occur in an antibody-accessible region (‘accessibility’ term); and (3) the mutation will disrupt antibody binding (‘dissimilarity’ term). All three terms can be computed from pre-pandemic data sources, providing early warning time critical for vaccine development. The accessibility and dissimilarity terms are computed using biophysical information. The fitness term is computed via the deep learning of evolutionary sequences. In particular, the authors computed the fitness term using EVE~\cite{frazer2021disease}, a deep generative model (i.e., VAE) trained on evolutionarily related protein sequences that learn constraints underpinning structure and function for a given protein family. 

Long-term and system-level evolution has also been systematically examined. Konno et al. clearly demonstrated that the evolution of gene content in metabolic systems is largely predictable by using ancestral gene content reconstruction and machine learning techniques~\cite{konno2023machine}. They first inferred the gene content of the ancestral species using the genomes of 2894 bacterial species (encompassing 50 phyla) and a reference phylogeny. Then they applied two classical machine learning models (logistic regression and random forest) to predict which genes will be gained or lost in metabolic pathway evolution, using the gene content vector of the parental node in the phylogenetic tree.  Their framework, Evodictor, successfully predicted gene gain and loss evolution at the branches of the reference phylogenetic tree, suggesting that evolutionary pressures and constraints on metabolic systems are universally shared. It would be interesting to see if deep learning techniques can be applied to predict metabolic system evolution.

\subsection{Microbe-X Interactions}

Recent advancements in microbiology and microbiome research have significantly deepened our understanding of the complex microbial interactions, as well as interactions between the microbes and the host, diseases, and drugs. In this section, we will discuss how deep learning-based methods have facilitated the inference of those complex interactions. 

\phantomsection
\subsubsection{Microbe-microbe interactions}
Microbes interact with each other and influence each other’s growth in various ways. Inferring the microbial interactions is important to understand the systems-level properties of the microbial communities. Typically, this is achieved by analyzing high-quality longitudinal~\cite{cao2017inferring,gerber2012inferring,stein2013ecological,steinway2015inference,bucci2016mdsine}, or steady-state data~\cite{xiao2017mapping}, which is hard to obtain for large-scale microbial communities. Recently, the traditional random forest classifier was proposed to tackle this issue~\cite{dimucci2018machine}. For each species, a trait is represented as a binary code in its trait vector. For each species pair within a community, a composite trait vector is created by concatenating the trait vectors of both species. This composite vector is then related to the observed responses of the interacting species. All interactions observed are utilized to train the classifier, which predicts the results of unobserved interactions. This approach has been evaluated in three case studies: a mapped interaction network of auxotrophic \textit{Escherichia coli} strains, a soil microbial community, and a comprehensive \textit{in silico} network illustrating metabolic interdependencies among 100 human gut bacteria. The results demonstrated that having partial knowledge of a microbial interaction network, combined with trait-level data of individual microbial species, can lead to accurate predictions of missing connections within the network, as well as propose potential mechanisms for these interactions. It would be very interesting to explore if deep learing methods can further improve the prediction of microbial interactions.

\phantomsection
\subsubsection{Microbe-host interactions}
A disrupted gut microbiome has been linked to a wide variety of diseases, yet the mechanisms by which these microbes affect human health remain largely unclear. Protein-protein interactions (PPIs) are increasingly recognized as a key mechanism through which gut microbiota influence their human hosts~\cite{lim2022artificial,post2022structural,balint2024human,pan2024microbial}. A vast and largely unexplored network of microbe-host PPIs may play a significant role in both the prevention and progression of various diseases. Future research is needed to uncover these interactions and their potential therapeutic implications.

Many machine learning methods have been developed to predict PPIs. Basically, they can be grouped into three categories: sequence-based, structure-based, and network-based~\cite{lim2022artificial}. Sequence-based methods utilize amino acid sequences to predict PPIs. For instance, PIPR employs a deep residual recurrent CNN within a siamese architecture to select local features and maintain contextual information without predefined features~\cite{chen2019multifaceted}.
Similarly, DeepPPISP integrates global and contextual sequence features by applying a sliding window approach to neighboring amino acids and utilizing a TextCNN architecture to treat the protein sequence as a one-dimensional image for global feature extraction~\cite{zeng2020protein}. Additionally, hybrid approaches have been developed for microbe-host PPI prediction, combining a denoising autoencoder (unsupervised learning) with logistic regression (supervised learning)~\cite{chen2020framework}. Another model, DeepViral, enhances performance by incorporating infectious disease phenotypes alongside protein sequences for microbe-host PPI prediction~\cite{liu2021deepviral}.

Structure-based methods leverage the three-dimensional structures of proteins to predict PPIs. For example, DeepInterface is one of the first methods to use 3D CNNs for predicting PPI interfaces at the atomic level~\cite{balci2019deepinterface}.
Different from DeepInterface, MaSIF (Molecular Surface Interaction Fingerprints) uses geometric deep learning to process non-Euclidean data, breaking proteins into overlapping patches with specific physicochemical properties to predict PPI interfaces~\cite{gainza2020deciphering}. Graph-based neural network methods, where nodes represent atoms or amino acid residues linked by edges based on spatial proximity or chemical bonds, apply convolutional filters on the graph representation of proteins to predict interactions while being invariant to rotation and translation. PECAN further integrates a graph CNN with an attention mechanism and transfer learning, using sequence-based conservation profiles and spatial distance features to predict antigen-antibody interactions~\cite{pittala2020learning}.

Network-based methods consider the PPI prediction problem as a link prediction task, using inferring missing links based on existing network knowledge. These methods have been benchmarked across various interactomes, demonstrating that advanced similarity-based methods, which leverage the network characteristics of PPIs, outperform other link prediction methods~\cite{wang2023assessment}. These general-purpose methods can be tailored for microbe-host PPI prediction. Moreover, integrating sequence-based, structure-based, and network-based approaches can leverage the strengths of each approach, potentially leading to more accurate and robust PPI predictions.

Of course, the microbe-host interactions are not limited to PPIs. Besides PPIs, microbes can interact with the host through many other mechanisms, including: (1) Gene regulation: Microbial metabolites can influence host gene expression via epigenetic changes or signaling pathways. (2) Immune modulation: Microbes interact with the host immune system, educating immune cells and promoting tolerance or inflammation. (3) Metabolite production: Gut microbes produce metabolites like short-chain fatty acids (SCFAs), which influence host energy metabolism, immune function, and gut health. (4) Gut barrier function: Microbes can strengthen or disrupt the gut barrier, affecting intestinal permeability.

\phantomsection
\subsubsection{Microbe-disease associations} 
The exploration of microbe-disease associations (MDAs) is crucial for understanding various health conditions and tailoring effective treatments. 
Traditional studies directly correlate microbial features with disease outcomes, creating MDA databases such as HMDAD \cite{ma2017analysis} and mBodyMap \cite{jin2022mbodymap}.
Advanced deep-learning methods have also been developed to infer new MDAs. Those methods include NinimHMDA~\cite{ma2020ninimhmda}, LGRSH~\cite{lei2020predicting}, BPNNHMDA~\cite{li2020identifying},
\newline
DMFMDA~\cite{liu2020dmfmda}, GATMDA~\cite{long2021predicting}, MVGCNMDA~\cite{hua2022mvgcnmda}, KGNMDA~\cite{jiang2022kgnmda}, GPUDMDA~\cite{peng2023predicting} and GCATCMDA~\cite{jiang2024predicting}.

NinimHMDA uses a multiplex heterogeneous network constructed from HMDAD and other biological databases~\cite{ma2020ninimhmda}. By integrating biological knowledge of microbes and diseases represented by various similarity networks and utilizing an end-to-end GCN-based mining model, it predicts different types of HMDAs (elevated or reduced) through a one-time model training. Predicting HMDAs is akin to solving a link-prediction problem within a multiplex heterogeneous network. In terms of predictive performance, NinimHMDA was compared with several existing methods such as DeepWalk \cite{perozzi2014deepwalk}, metapath2vec \cite{dong2017metapath2vec}.

Similar to NinimHMDA, LGRSH~\cite{lei2020predicting} and BPNNHMDA~\cite{li2020identifying} were developed for the same predictive task but with different deep-learning architectures. LGRSH applies graph representation techniques to predict associations, using calculated similarities between microbes and diseases~\cite{lei2020predicting}. BPNNHMDA uses a back-propagation neural network to predict potential associations~\cite{li2020identifying}. DMFMDA employs deep matrix factorization and Bayesian Personalized Ranking to predict associations~\cite{liu2020dmfmda}.

Thanks to the advancements in large language models, extraction of MDAs directly from biomedical literature has become much easier than before. For example, Karkera et al. demonstrated that pre-trained language models (specifically GPT-3, BioMedLM, and BioLinkBERT), when fine-tuned with domain and problem-specific data, can achieve state-of-the-art results for extracting MDAs from scientific publications~\cite{karkera2023leveraging}. The extracted MDAs will further expand the human MDA database. We expect that those deep learning methods will be more powerful with an expanded human MDA database.

Deep learning techniques have also been leveraged to study the association between microbes and specific diseases. For instance, MICAH is a deep learning method based on a heterogeneous graph transformer to study the relationships between intratumoral microbes and cancer tissues~\cite{liu2023explainable}.  The inputs of MICAH are the species abundance matrix and sample labels (i.e., cancer types of samples). From the inputs, MICAH constructs a heterogeneous group with two types of nodes (microbes and samples), and three types of edges (species-species metabolic edges based on the NJS16 database~\cite{sung2017global}, species-species phylogenetic edges based on the NCBI Taxonomy database, species-sample edges representing the relative abundance of a species in a sample). Then, MICAH used a two-layer graph transformer to update node embeddings and a fully connected layer based on updated node embeddings to perform sample node (cancer type) classification. Finally, MICAH extracts the attention scores of species to samples from the well-trained model to output subsets of microbial species associated with different cancer types. This framework significantly refines the number of microbes that can be used for follow-up experimental validation, facilitating the study of the relationship between tumors and intratumoral microbiomes.

\phantomsection
\subsubsection{Microbe-drug associations} 
Accumulated clinical studies show that microbes living in humans interact closely with human hosts, and get involved in modulating drug efficacy and drug toxicity. Microbes have become novel targets for the development of antibacterial agents. Therefore, screening of microbe–drug associations can benefit greatly drug research and development. With the increase of microbial genomic and pharmacological datasets, we are greatly motivated to develop effective computational methods to identify new microbe–drug associations.

Many deep-learning methods have been recently developed to identify microbe–drug associations, e.g., GARFMDA~\cite{kuang2024novel}, GCNATMDA~\cite{wang2024microbe}, LCASPMDA~\cite{yang2024lcaspmda}, MCHAN~\cite{li2024mchan}, \\
MDSVDNV~\cite{tan2024mdsvdnv}, NMGMDA~\cite{liang2024nmgmda}, OGNNMDA~\cite{zhao2024ognnmda}, STNMDA~\cite{Liu2024stnmda}, etc. Most of the deep learning methods can be divided into six different categories based on the deep learning model they used~\cite{wang2022review}, e.g., CNN-based, GCN-based autoencoder, Graph Attention Network(GAT)-based autoencoder, Collective Variational Autoencoder (CVAE), Sparse Autoencoder (SAE). A recent method STNMDA is an exception~\cite{fan2024stnmda}. STNMDA integrates a Structure-Aware Transformer (SAT) with an MLP classifier to infer microbe-drug associations. It begins with a ``random walk with a restart'' approach to construct a heterogeneous network using Gaussian kernel similarity and functional similarity measures for microorganisms and drugs. This heterogeneous network was then fed into the SAT to extract attribute features and graph structures for each drug and microbe node. Finally, the MLP classifier calculated the probability of associations between microbes and drugs. A systematic comparison of those existing methods using benchmark datasets is warranted.

\subsection{Microbial Ecology}
Microbial ecology seeks to decode the intricate dynamics and functional roles within complex microbial communities, from soil rhizospheres to the human gut. AI has emerged as a transformative force in this domain, enabling the extraction of predictive patterns from high-dimensional, noisy, and often sparse datasets that defy traditional statistical modeling. This section explores key applications where AI drives breakthroughs in microbial ecology: from predicting community composition to identifying keystone species, predicting community dynamics, community control, and source tracking. By integrating AI techniques with ecological theory, these applications not only enhance mechanistic understanding but also empower predictive stewardship of microbiomes in natural and engineered systems.


\phantomsection
\subsubsection{Microbial composition prediction}
Being able to predict microbial community composition is a cornerstone of modern microbiome science. Microbial assemblages underpin essential ecological and physiological processes—from nutrient cycling in the environment to metabolism and immunity in host-associated systems—and their composition strongly determines overall community function. Predictive capability marks a shift from descriptive cataloging (``who is there") toward a mechanistic understanding of why and how communities assemble and change. Such models provide insight into the deterministic and stochastic forces shaping microbial ecosystems and enable the rational design and control of microbiomes for desired outcomes, including bioremediation, agricultural productivity, and human health interventions. In medicine, for example, accurate prediction of community compositions could inform diet-based therapies, probiotic design, and microbiota restoration after antibiotic treatment. Ultimately, the ability to forecast microbial community composition transforms microbiome research into a predictive, intervention-oriented discipline, akin to how meteorology evolved from observation to accurate weather forecasting.

Early attempts include a bioclimatic modeling approach that leverages neural networks to predict microbial community structure as a function of environmental parameters and microbial interactions~\cite{larsen2012predicting}, and the use of MLP to predict the temporal gut community composition  of termite perturbed by six different lignocellulose food sources~\cite{benjamino2018low}.

cNODE (compositional neural ordinary differential equation) is a deep learning method that can predict the community compositions from the species assemblages for a given ecological habitat of interest, e.g., the human gut~\cite{michelmata_predicting_2022}. All microbial species that can inhabit this habitat form a species pool or meta-community. A microbiome sample collected from this habitat can be considered as a local community assembled from the meta-community. The species assemblage of this sample is characterized by a binary vector, where the entry indicates if species-$i$ is present (or absent) in this sample. The community composition is characterized by a compositional vector, where the $i$th-entry represents the relative abundance of species-$i$. cNODE aims to implicitly learn the community assembly rules by learning the mapping from species assemblage into community composition. To learn such a mapping, cNODE used Neural ODE~\cite{chen_neural_2018}, which can be interpreted as a continuous limit of the ResNet architecture~\cite{he_deep_2016}. Extensive simulations suggest that the sample size in the training data acquired to reach a relatively accurate prediction should be twice the species pool size. cNODE has been successfully applied to predict compositions of the ocean and soil microbiota, Drosophila melanogaster gut microbiota, and the human gut and oral microbiota.

Instead of relying on species assemblage, MicrobeGNN employs a graph neural network-based approach to predict the microbial composition at steady state from the genomes of mixed bacteria, with each species represented by a node~\cite{ruaud2024modelling}. Bacterial genomes are encoded into binary feature vectors that indicate the presence or absence of specific genes. Two types of GNNs, GraphSAGE~\cite{hamilton2017inductive} and MPGNN~\cite{kipf_semi-supervised_2017}, are utilized for node and edge computations, respectively. Due to the lack of prior knowledge regarding the exact graph topology, fully connected graphs are employed, allowing each node to influence all other nodes within a single message-passing step. The results demonstrate that GNNs can accurately predict the relative abundances of bacteria in communities based on their genomes across various compositions and sizes.

Note that neither cNODE nor MicrobeGNN utilizes environmental or host factors in predicting microbial compositions. Incorporating environmental/host factors into deep learning models might further improve the accuracy of microbial composition predictions.

\phantomsection
\subsubsection{Keystone species identification}
Keystone species are species whose influences on community structure and functioning are disproportionately large relative to its abundance. Microbial communities are also thought to contain keystone species, but applying macro-ecological approaches to large, complex microbiomes is challenging. Experimental removals are infeasible due to technical limitations and ethical concerns, particularly in host-associated systems such as the human gut microbiome. Statistical comparisons are similarly difficult, as it is rare to find two microbial communities differing by only a single species, and such comparisons are further complicated by strong inter-individual variation and numerous confounders. An alternative is to infer population-dynamics models and identify keystone species through simulated removals, but these methods require accurate model specification and high-quality absolute abundance data. Keystone species were also considered highly connected or high-betweenness taxa in microbial correlation networks. However, correlation-based edges capture statistical co-occurrence rather than direct ecological interactions, and network topology does not account for community specificity, e.g., a species may function as a keystone in one community but not in another.

By implicitly learning the community assembly rules, cNODE or its variant enables us to predict the new community compositions after adding or removing any species or any species combinations via thought experiments. In particular, predicting the impact of species’ removal facilitates the identification of keystone species that have a disproportionately large effect on the structure or function of their community relative to their abundance~\cite{wang2024identifying}. 
Note that the impact of a species’ removal naturally depends on the resident community, i.e., a species may be a keystone in one community but not necessarily a keystone in another community. In other words, the keystoneness of a species can be highly community-specific.

The DKI (Data-driven Keystone species Identification) framework is based on cNODE~\cite{wang2024identifying}. In the DKI framework, the keystoneness of species in microbial communities was defined as the product of two components: the impact component and the biomass component. The impact component quantifies the impact of species's removal on the structure of community, while the biomass component captures how disproportionate this impact is. 

The DKI framework was validated using synthetic data generated from a classical population dynamics model in community ecology, i.e., the Generalized Lotka-Volterra (GLV) model, and then applied to compute the keystoneness of species in the human gut, oral microbiome, and the soil and coral microbiome. It was found that those taxa with high median keystoneness across different samples display strong community specificity, and some of them have been reported as keystone taxa in literature. Instead of studying the impact of removing a single species, the DKI framework can be extended to study the impact of removing any species combinations,  and hence study keystone duos or trios, etc, in complex microbial communities. Instead of studying the impact of removing a single species, the DKI framework can be extended to study the impact of removing any species combinations,  and hence study keystone duos or trios, etc, in complex microbial communities.

\phantomsection
\subsubsection{Microbial dynamics prediction}
A fundamental question in microbial ecology is whether we can predict the temporal behaviors of complex microbial communities~\cite{chen2024stability}. Traditionally, this problem is addressed using system identification or network reconstruction techniques, which assume specific population dynamics described by a set of ordinary differential equations. For example, the classical GLV model in community ecology, which considers pair-wise interactions, can be represented as a directed, signed, and weighted graph, often referred to as an ecological network. Numerous methods have been developed to infer these dynamics and reconstruct the ecological network using temporal or steady-state data~\cite{liu2023controlling}. However, this network-based approach typically assumes that inter-species interactions are exclusively pair-wise, which may not reflect the true nature of complex microbial interactions.

Recently, deep learning techniques have been deployed to predict temporal behaviors of microbiomes. For example, in 2022, Baranwal et al. applied LSTM (a classical variant of RNN) to learn from experimental data on temporal dynamics and functions of microbial communities to predict their future behavior and design new communities with desired functions~\cite{baranwal_recurrent_2022}. Using a significant amount of experimental data, they found that this method outperforms the widely used GLV model in community ecology. In 2023, Thompson et al. proposed the Microbiome Recurrent Neural Network (MiRNN) architecture~\cite{thompson2023integrating}. Inputs to the MiRNN at time step $t-1$ include the state of species abundances, metabolite concentrations, control inputs, and a latent vector that stores information from previous steps and whose dimension determines the flexibility of the model.The output from each MiRNN block is the predicted system state and the latent vector at the next time step $t$. To avoid the physically unrealistic emergence of previously absent species, a constrained feed-forward neural network outputs zero-valued species abundances if species abundances at the previous time step were zero. The authors demonstrated that MiRNN yielded comparable prediction performance to the LSTM model, but with more than a 50,000 fold reduction in the number of model parameters. These works are of broad interest to those working on microbiome prediction and design to optimize specific target functions. So far, LSTM and MiRNN have been just applied to synthetic communities with 25 diverse and prevalent human gut species and 4 major health-relevant metabolites (acetate, butyrate, lactate, and succinate). Its potential to large systems, e.g., the human gut microbiome, with thousands of species and metabolites would be interesting to explore. The quality of the training data would be crucial.

Autoencoders (AEs) have recently been employed to compress microbial growth curves into compact, low-dimensional embeddings~\cite{baig2023autoencoder}. These embeddings enable high-fidelity reconstruction of the original growth curves while preserving critical biological information despite substantial dimensionality reduction. Remarkably, they retain sufficient signal to distinguish complex phenotypes—such as strain identity and antibiotic resistance—predict microbial dynamics from initial conditions and experimental variables, and map directly to interpretable mechanistic growth parameters. In many cases, AE-derived embeddings outperform full growth curves on these tasks, indicating effective removal of extraneous noise without sacrificing biologically relevant information. Furthermore, AEs facilitate the compression of community dynamics into fewer variables than typically required by conventional mechanistic models.

In addition to methods specifically designed for predicting microbial dynamics, existing methodologies developed for multiple time series forecasting (MTSF) can also be potentially employed. For example, MTSF-DG is a model capable of learning historical relation graphs and predicting future relation graphs to capture dynamic correlations~\cite{zhao2023multiple}. Evaluating the performance of these general time series prediction methods in the context of microbial dynamics prediction would be very interesting.

Predicting the colonization outcomes of exogenous species for complex microbial communities can be considered as a special dynamics prediction problem~\cite{wu2024data}. Machine learning methods treat the baseline (i.e., pre-invasion) taxonomic profile as inputs and the steady state abundance of the invasive species as output or mathematically, learn the mapping from the baseline taxonomic profile of a community to the steady state abundance of the invading species.
Validation of the approach using synthetic data and two commensal gut bacteria species \textit{Enterococcus faecium} and \textit{Akkermansia muciniphila} in hundreds of human stool-derived \textit{in vitro} microbial communities, showed that machine learning models, including random forest, linear regression/logistic regression, and neural ODE can predict not only the binary colonization outcome but also the final abundance of the invading species~\cite{wu2024data}.

Fecal microbiota transplantation (FMT) has shown a high success rate for the treatment of recurrent \textit{Clostridioides difficile} infection (rCDI). However, the mechanisms and dynamics dictating which donor microbiomes can engraft in the recipient are poorly understood. Traditional machine learning models, e.g., random forest, have been applied to predict the post-FMT bacterial species engraftment~\cite{ianiro_variability_2022}.
We expect that, given high-quality training data, deep learning methods can also be used to predict species engraftment and outperform traditional machine learning methods.  

\phantomsection
\subsubsection{Control of microbial communities}
Multi-species microbial communities are ubiquitous in natural ecosystems. When engineered for biomanufacturing, synthetic co-cultures outperform monocultures by enhancing productivity and reducing metabolic burden through task compartmentalization across subpopulations. Despite these advantages, precise control of multi-species communities in bioreactors remains a major challenge—the primary reason most industrial bioprocessing still relies on monocultures. Recently, deep reinforcement learning (DRL)—reinforcement learning augmented with deep neural networks—has been successfully applied to regulate interacting microbial populations in bioreactors~\cite{treloar2020deep}. Reinforcement learning helps agents learn decision-making through trial and error. DRL improves this by using deep learning to extract decisions from unstructured data without manual state space engineering. DRL algorithms can take in very large inputs (e.g., an image of the raw board state and the history of states) and decide what actions to perform to optimize an objective (e.g., winning the game). A famous DRL algorithm is AlphaGo Zero, learning from playing the ancient Chinese game of Go without using any human knowledge~\cite{silver2017mastering}. To control the microbial populations in a bioreactor, feedback from a trained DRL agent effectively maintains species at target abundances. Notably, model-free DRL with bang-bang control outperforms traditional proportional-integral controllers under infrequent sampling conditions. Moreover, a robust control policy can be learned in a single 24-hour experiment using just five parallel bioreactors. Finally, DRL can directly optimize bioprocess yield in co-cultures. Crucially, this model-free approach requires no prior knowledge of population interactions, making it broadly applicable. We anticipate that DRL will play a pivotal role in microbial community control.

\phantomsection
\subsubsection{Microbiome data simulation}
In microbiome research, we often need to generate synthetic data for testing computational methods, augment existing data to enhance model training, or impute missing data points for downstream data analyses. There are two primary approaches to achieve these goals: (1) statistical or population dynamics models; (2) deep generative models.

CAMISIM~\cite{fritz2019camisim} is a microbial community and metagenome simulator. It can model different microbial abundance profiles, multi-sample time series, and differential abundance studies, includes real and simulated strain-level diversity, and generates second- and third-generation sequencing data from taxonomic profiles or de novo. CAMISIM consists of three steps: (1) Design of the microbial community, which includes selection of the community members and their genomes, and assigning them relative abundances; (2) Metagenome sequencing data simulation; and (3) Postprocessing, where the binning and assembly gold standards are produced. 
TADA~\cite{sayyari2019tada} uses available microbiome data and a statistical generative model to create new samples distributed around existing samples, taking into account phylogenetic relationships between microbial species. TADA models two types of variations: \emph{true variation} due to confounding factors or natural biological variation among samples; \emph{sampling variation} due to the fact that sequencing takes a random (but not necessarily uniformly random) subsample of the true diversity, creating additional variation around the true proportions. 
SparseDOSSA~\cite{ma2021statistical} is a statistical model designed to represent microbial ecological population structures . It models marginal microbial feature abundances using a zero-inflated lognormal distribution, incorporating additional components for absolute cell counts, the sequence read generation process, and interactions between microbes and their environment.
AugCoDa~\cite{gordon2022data} combines key principles from compositional data analysis (e.g., the Aitchison geometry of the simplex and subcompositions) with classical data augmentation techniques (e.g., Mixup~\cite{zhang2018mixup} and CutMix~\cite{yun2019cutmix}).  
PhyloMix~\cite{jiang2025phylomix} leverages the phylogenetic relationships among microbiome taxa as an informative prior to guide the generation of synthetic microbial samples. Leveraging phylogeny, PhyloMix creates new samples by removing a subtree from one sample and combining it with the corresponding subtree from another sample. 
MIDASim~\cite{he2024midasim} is a fast and simple approach for simulating realistic microbiome data that reproduces the distributional and correlation structure of a template microbiome dataset. MIDASim consists of two steps. The first step generates correlated binary indicators that represent the presence-absence status of all taxa, and the second step generates relative abundances and counts for the taxa that are considered to be present in step 1, utilizing a Gaussian copula to account for the taxon-taxon correlations.

miaSim~\cite{gao2023miasim} is a versatile R/Bioconductor package, providing tools to simulate (longitudinal) time series data from popular models in microbial ecology, e.g., Self-organised instability (SOI), Hubbell's neutral model, generalized Lotka-Volterra (gLV), Ricker model (discrete gLV), Stochastic logistic model, MacArthur’s consumer-resource model (CRM). 
MIMIC~\cite{fontanarrosa2025mimic} is a Python package designed to advance the simulation, inference, and prediction
of microbial community interactions and dynamics. MIMIC integrates a suite of mathematical models, including gLV, Gaussian Processes (GP), and Vector Autoregression (VAR) to offer a versatile framework for analyzing microbial dynamics.

Generative deep learning techniques, such as GAN and AE, can be naturally employed for microbiome data generatation/imputation. 
For example, MB-GAN~\cite{rong_mb-gan_2021} learns latent spaces from observed microbial abundances and generates simulated abundances based on these learned distributions. By modifying the discriminator network in the original GAN framework to incorporate microbiome diversity-based measurements, MB-GAN captures visual patterns of sequencing profiles and generates realistic human gut microbiome profiles. 
DeepBioGen\cite{choi_deepmicrogen_2023} captures local visual patterns of sequencing profiles by training conditional Wasserstein GAN, whose generator and critic networks are composed of up-convolutional and convolutional layers, respectively. DeepBioGen can then generates new sequencing profiles capturing those local visual patterns. 
phylaGAN~\cite{sharma2024phylagan} is a deep learning framework to augment the existing datasets with generated microbiome data using a combination of conditional GAN and autoencoder. The conditional GAN trains two models against each other to compute larger simulated datasets that are representative of the original dataset, while the autoencoder maps the original and the generated samples onto a common subspace to make the prediction more accurate. 
mbSparse~\cite{qi2025mbsparse} leverages a feature autoencoder (AE) to learn sample representations and a conditional VAE for data reconstruction. The feature AE generates representative embeddings from the input matrix (e.g., the raw OTU matrix) using a stacked architecture of two dense networks in both the encoder and decoder stages. The conditional VAE enhances the traditional VAE by incorporating conditional variables, enabling the generation of diverse, targeted data tailored to specific conditions. 
Note that the methods were originally designed for data augmentation of cross-sectional microbiome data. For longitudinal microbiome data imputation, DeepMicroGen~\cite{choi_deepmicrogen_2023} offers a robust solution. It extracts
features that incorporate phylogenetic relationships between taxa using CNN. These features are subsequently processed by a bidirectional RNN-based GAN model, which generates imputed values by learning the temporal dependencies between observations at different time points.

These approaches enhance our ability to generate high-fidelity synthetic microbiome data, crucial for developing and testing new methods for microbiome data analysis.

\phantomsection
\subsubsection{Microbial source tracking}
Determining the contributions of various environmental sources (``sources'') to a specific microbial community (``sink'') represents a traditional challenge in microbiology, commonly referred to as microbial source tracking (MST). Addressing this MST challenge will not only enhance our understanding of microbial community formation but also has significant implications in areas like pollution management, public health, and forensics. MST techniques are generally categorized into two types: target-based methods, which concentrate on identifying source-specific indicator species or chemicals, and community-based methods, which analyze community structures to assess the similarity between sink samples and potential source environments. With next-generation sequencing becoming standard for community assessment in microbiology, numerous community-based computational methods, known as MST solvers, have been developed and applied to various real-world datasets, showcasing their effectiveness across different scenarios.

Here, we introduce some representative MST solvers. The first solver is based on the classification analysis in machine learning, for example, using the random forest classifier. In this case, each source represents a distinct class, and the classifier will classify the sink into different classes with different probabilities. The probabilities of the sink belonging to the different classes can be naturally interpreted as the mixing proportions or contributions of those sources to the sink. Beyond the simple classification analysis, more advanced statistical methods based on Bayesian modeling have been developed. For example, SourceTracker is a Bayesian MST solver that explicitly models the sink as a convex mixture of sources and infers the mixing proportions via Gibbs sampling~\cite{knights2011bayesian}. FEAST (fast expectation-maximization for microbial source tracking~\cite{shenhav2019feast}) is a more recent statistical method. FEAST also assumes each sink is a convex combination of sources. But it infers the model parameters via fast expectation-maximization, which is much more scalable than Markov Chain Monte Carlo used by SourceTracker. STENSL (microbial Source Tracking with ENvironment SeLection) is also based on expectation-maximization~\cite{an2022stensl}. STENSL enhances traditional MST analysis through unsupervised source selection and facilitates the sparse identification of hidden source environments. By integrating sparsity into the estimation of potential source environments, it boosts the accuracy of true source contributions and considerably diminishes the noise from non-contributing sources. ONN4MST is a deep learning method based on the Ontology-aware Neural Network (ONN) to solve large-scale MST problems~\cite{zha_ontology-aware_2022}. The ONN model promotes predictions in line with the ``biome ontology.'' Essentially, it leverages biome ontology information to represent the relationships among biomes and to estimate the distribution of different biomes within a community sample. The authors demonstrated clear evidence that ONN4MST outperformed other methods (e.g., SourceTracker and FEAST) with near-optimal accuracy when source tracking among 125,823 samples from 114 niches. 

Many MST solvers draw inspiration from the analogy between the MST problem and estimating the mixing proportions of conversation topics in a test document. It has been pointed out that this analogy is problematic~\cite{wang2023ecological}. In topic modeling~\cite{griffiths2004finding}, a specialized area within NLP, the objective is to uncover the abstract ``topics'' present in a set of documents, which can be viewed as static or ``dead.'' In contrast, MST typically involves dynamic, thriving microbial communities where ecological dynamics significantly influence community assembly and their state, that is, the microbial composition. Given these ecological dynamics, a sink community cannot merely be viewed as a convex mixture of known and unknown sources. Indeed, through numerical simulations, analytical calculations, and real data analysis, compelling evidence has been presented that ecological dynamics impose fundamental challenges in community‐based MST~\cite{wang2023ecological}. Thus, results from current MST solvers require very cautious interpretation.

\subsection{Metabolic Modeling}

Metabolic modeling has become a crucial component in microbiology and microbiome research, significantly enhancing our understanding of microbial interactions and their effects on environments or host well-being. This approach integrates computational methods with biological insights, facilitating the prediction, analysis, and comprehension of metabolic capabilities and interactions within microbial communities.

\phantomsection
\subsubsection{Gap filling: inferring missing reactions } 
Genome-scale metabolic models (GEMs) have substantially advanced our understanding of the complex interactions among genes, reactions, and metabolites. These models, integrated with high-throughput data, support applications in metabolic engineering and drug discovery. For instance, AGORA2 (Assembly of Gut Organisms through Reconstruction and Analysis, version 2), representing the cutting-edge GEM resource for human gut microorganisms, comprises 7,302 strains and provides strain-resolved capabilities for drug degradation and biotransformation for 98 drugs~\cite{heinken2023genome}. This resource has been meticulously curated using comparative genomics and extensive literature reviews. AGORA2 facilitates personalized, strain-resolved modeling by predicting how patients’ gut microbiomes convert drugs. Additionally, AGORA2 acts as a comprehensive knowledge base for the human microbiome, paving the way for personalized and predictive analyses of host–microbiome metabolic interactions. Reconstruction of GEMs typically require extensive manual curation to improve their quality for effective use in biomedical applications. Yet, due to our imperfect knowledge of metabolic processes, even highly curated GEMs could have knowledge gaps (e.g., missing reactions). Various optimization-based gap-filling methods have been developed to identify missing reactions in draft GEMs \cite{orth2010systematizing, pan2018advances, rana2020recent}. 

The existing gap-filling methods often require experimental data, but such experimental data is scarce for non-model organisms, limiting tool utility. If not using any domain knowledge, gap-filling of GEMs or inferring missing reactions in GEMs purely from the topology of the GEM can be treated as a hyperlink prediction problem~\cite{chen2023survey}. As we know, we can always consider a metabolic network or any biochemical reaction network as a hypergraph, where metabolites are nodes, reactions are hyperlinks.  For instance, Chen et al. present the Chebyshev spectral hyperlink predictor (CHESHIRE), a deep learning-based method for identifying missing reactions in GEMs based on the topology of metabolic networks \cite{chen2023teasing}. CHISHIRE leverages the Chebyshev spectral GCN on the decomposed graph of a metabolic network to refine the feature vector of each metabolite by incorporating the features of other metabolites from the same reaction. As a variant of GCN, Chebyshev spectral GCN was designed to efficiently process data represented as graphs~\cite{defferrard2016convolutional}. It leverages spectral graph theory and Chebyshev polynomials to perform graph convolutions in the spectral domain. It has been shown that CHESHIRE outperforms other topology-based hyperlink rediction methods, e.g., Neural Hyperlink Predictor (NHP)~\cite{yadati2020nhp} and C3MM Clique Closure-based Coordinated Matrix Minimization (C3MM)~\cite{sharma2021c3mm} in predicting artificially removed reactions over 926 GEMs (including 818 GEMs from AGORA). Furthermore, CHESHIRE is able to improve the phenotypic predictions of 49 draft GEMs for fermentation products and amino acids secretions. Both types of validation suggest that CHESHIRE is a powerful tool for GEM curation..


\phantomsection
\subsubsection{Retrosynthesis: breaking down a target molecule}
Note that gap-filling is the strategy used to complete metabolic networks when certain reactions or pathways are missing. It identifies reactions that need to be added to a metabolic model to ensure the system can produce all required metabolites and metabolic phenotypes. Retrosynthesis is a complementary strategy. Retrosynthesis involves iteratively breaking down a target molecule into simpler molecules that can be combined chemically or enzymatically to produce it. Eventually, all the required compounds are either commercially available or present in the microbial strain of choice. Retrosynthesis is used to map out potential biosynthetic pathways to produce a desired compound by analyzing reaction steps in reverse. While gap-filling aims to ensure the completeness of the metabolic network for overall functionality, retrosynthesis focuses on pathway construction for a specific product. Recently, a reinforcement learning method RetroPath RL was developed for bioretrosynthesis~\cite{koch2019reinforcement}. RetroPath RL is based on the Monte Carlo Tree Search (MCTS), which is a heuristic search algorithm combining the principles of random sampling (Monte Carlo methods) and search trees to balance exploration and exploitation in making optimal decisions~\cite{coulom2006efficient,silver2016mastering}. RetroPath RL takes as input a compound of interest, a microbial strain as a sink (i.e., the list of available precursor metabolites) and a set of reaction rules, e.g., RetroRules, a database of reaction rules for metabolic engineering~\cite{duigou2019retrorules}.

One interesting application of RetroPath RL is to complete further the metabolism of specific compounds in the human gut microbiota. For instance, Balzerani et al. used RetroPath RL to predict the degradation pathways of phenolic compounds~\cite{balzerani2022prediction}. By leveraging Phenol-Explorer~\cite{rothwell2013phenol}, the largest database of phenolic compounds in the literature, and AGREDA~\cite{blasco2021extended}, an extended metabolic network amenable to analyze the interaction of the human gut microbiota with diet, the authors generated a more complete version of the human gut microbiota metabolic network.

\phantomsection
\subsubsection{Modeling microbial metabolism}
Metabolic modeling has long provided critical insights into microbial community metabolism. However, conventional flux balance analysis (FBA) methods typically fail to predict adaptive metabolic and regulatory strategies that promote long-term survival and stability, particularly in heterogeneous communities. Recently, SPAM-DFBA (Self-Playing Microbes in Dynamic FBA)~\cite{ghadermazi2024microbial} was introduced—an RL framework that reimagines microbial metabolism as a decision-making process. In SPAM-DFBA, each GEM is treated as an agent that learns to regulate metabolic fluxes based on observable environmental states. Through a self-play mechanism, agents evolve adaptive strategies via trial and error, without requiring predefined metabolic or regulatory rules. “Maladaptive” decisions are naturally filtered through simulated selection, enabling the discovery of policies that maximize long-term fitness.
Unlike traditional FBA and dynamic FBA (DFBA), which typically assume biomass maximization at a single time point, SPAM-DFBA accounts for the long-term consequences of metabolic decisions in dynamic, multi-species contexts. Greedy biomass optimization often leads to early community extinction—outcomes that RL-trained agents learn to avoid. The resulting flux regulation policies reveal why certain behaviors emerge in real microbial systems and predict stable metabolic strategies in complex, interacting ecosystems with minimal reliance on prior assumptions.

\subsection{Precision Nutrition}
Machine-learning models have shown remarkable accuracy in predicting metabolite profiles from microbial compositions~\cite{bar2020reference,reiman2021mimenet,wang2023predicting}. Furthermore, the intersection of computational biology with nutrition science has led to notable strides in personalized nutrition and food quality prediction~\cite{zeevi_personalized_2015,rein2022effects,wang2025predicting,ferreira2025assessing}. This emerging field focuses on customizing dietary recommendations to individual biological and physiological profiles, aiming to optimize health outcomes. By employing machine learning algorithms and microbiome data analysis, researchers are able to predict individual responses to various foods and diets, marking a significant advancement in the field of precision nutrition.

\phantomsection
\subsubsection{Nutrition profile correction}
An unhealthy diet is associated with higher risks of various diseases \cite{hu2002optimal, afshin2019health}. Measuring dietary intake in large cohort studies is often difficult, so we frequently depend on self-reported tools (like food frequency questionnaires, 24-hour recalls, and diet records) that are established in nutritional epidemiology~\cite{mcnutt2008development, sharpe2021automated, hebert1997development}. However, these self-reported instruments can be susceptible to measurement errors~\cite{westerterp2002validity}, resulting in inaccuracies in nutrient profile calculations. Although nutritional epidemiology uses methods such as regression calibrations \cite{rosner1989correction, spiegelman1997regression} and cumulative averages \cite{hu1999dietary} to address these inaccuracies, deep-learning approaches have not been leveraged to correct random measurement errors.

Wang et al. introduce a deep-learning method called METRIC (Microbiome-based Nutrient Profile Corrector) that utilizes gut microbial compositions to correct random measurement errors in nutrient profiles derived from self-reported dietary assessments \cite{wang2024microbiome}. METRIC draws inspiration from Noise2Noise, a deep learning model for image denoising in computer vision that reconstructs clean images using only corrupted inputs~\cite{Lehtinen2018Noise2Noise}. The core concept of Noise2Noise is training the model on pairs of noisy images as both the input and output, compelling the neural network to estimate the average of these corrupted images. This process leads the predictions to statistically align with the clean image due to the zero-mean property of the noise. In a similar way, METRIC addresses random errors in the assessed nutrient profile generated from self-reported dietary assessments, without using clean data (i.e., the ground truth dietary intake). It's important to note that METRIC targets the correction of the nutrient profile rather than the food profile (or the original dietary assessment), since the high frequency of zero values in the food profile—many food items not consumed—poses significant challenges for machine learning. In contrast, the derived nutrient profile tends to contain predominantly non-zero values. Additionally, METRIC aims to rectify random errors characterized by zero means, instead of systematic biases or errors with non-zero means, as correcting the latter effectively necessitates access to the ground truth dietary intake, which is often unavailable.

\phantomsection
\subsubsection{Metabolomic profile prediction}
Predicting the metabolomic profile (i.e., quantified amount of metabolites within a biological sample) from the composition of a microbial community is an active area in microbiome research. Experimental measurement of metabolites relies on expensive and complex techniques like Liquid Chromatography-Mass Spectrometry, which have incomplete coverage~\cite{letertre2020combined, alseekh2021mass}. In contrast, microbial composition measurements are cheaper, more automated, and have better coverage. Therefore, it is desirable to develop computational methods that predict metabolomic profiles based on microbial compositions~\cite{mallick2019predictive,reiman2021mimenet,wang2023predicting}. Additionally, such a method could facilitate our understanding of the interplay between microorganisms and their metabolites.

Various machine-learning methods have been developed to solve this problem. For example, MelonnPan uses an elastic net linear regression to model the relative abundance of each metabolite using metagenomic features~\cite{mallick2019predictive}. It simply models each metabolite individually, missing the opportunity to use shared information across metabolomic features to boost prediction performance. Similar to how word2vec~\cite{Mikolov2013Efficient} estimates word probabilities conditioned on a single particular word, mmvec~\cite{morton2019learning} takes microbial sequence counts and metabolite abundances from various samples as the input and outputs the estimated conditional probabilities of observing a metabolite given the presence of a specific microbe. These conditional probabilities generate the entire metabolite abundance profile for a given microbiome sample through a multinomial distribution. Note that in the original application of word2vec, the skip-gram technique (i.e., creating word embeddings that focus on predicting surrounding words based on a specific word or target word) was employed to account for the sequential nature of the text. For microbiome data, there is no clear sequential nature. Therefore, in mmvec, the skip-gram was replaced by multinomial sampling, where a single microbe is randomly sampled from a microbiome sample at each gradient descent step. Neural encoder-decoder (NED) leverages the constraints of sparsity and non-negative weights for mapping microbiomes to metabolomes~\cite{le2020deep}. The use of non-negative weights in NED imposes a stringent constraint on the model, which simplifies the model complexity but may limit the learning capacity. MiMeNet (Microbiome-Metabolome Network) is essentially an MLP that models the community metabolome profile using metagenomic taxonomic or functional features obtained from a microbiome sample~\cite{reiman2021mimenet}.



mNODE \cite{wang2023predicting} leverages a relatively new class of deep neural networks, i.e., neural ODE \cite{chen_neural_2018}, to predict metabolomic profiles from microbial compositions. Neural ODE can be considered the continuous-depth version of a Residual Network (ResNet). Compared with traditional MLPs, Neural ODEs model continuous-time dynamics, enabling smoother and more interpretable representations of temporal processes. Since the input dimension (the number of microbial species) is typically different from the output data (the number of microbial species), mNODE integrates the neural ODE as a middle module, sandwiched by two densely connected layers to adjust for data dimension variability. mNODE shows superior performance in both synthetic and real datasets than existing methods. Additionally, mNODE can naturally incorporate dietary information into its analysis of human gut microbiomes, further improving metabolomic profile predictions. Its susceptibility analysis can uncover microbe–metabolite interactions, which can be confirmed with both synthetic and real datasets.

\phantomsection
\subsubsection{Personalized diet recommendation} 
In recent years, the intersection of gut microbiome, nutrition science, and machine learning has led to significant advancements in personalized nutrition and food quality prediction. This emerging field aims to tailor dietary recommendations to individual biological and physiological factors (e.g., gut microbial composition), thereby optimizing health outcomes~\cite{zeevi_personalized_2015,salathe2024personalized,rein2022effects,wang2025predicting}.

Numerous studies use traditional machine learning methods to predict blood glucose levels based on the time-series data from continuous glucose monitor~\cite{li2016smartphone, cheng2024toward}. Similarly, Kim et al. apply RNN to predict blood glucose levels in hospitalized patients with type-2 diabetes \cite{kim2020developing}. Recently, Lutsker et al. present GluFormer, a generative foundation model based on the Transformer architecture to predict blood glucose measurements from non-diabetic individuals \cite{lutsker2024glucose}. However, these models do not incorporate dietary information in their inputs, limiting their ability to generate personalized dietary recommendations. In contrast, leveraging mathematical models and Bayesian statistics, Albers et al. predict an individual's postprandial blood glucose level using the preprandial blood glucose level and carbohydrate intake \cite{albers2017personalized}. 

Zeevi et al. use the gradient-boosting regressor (GBR) to predict personalized postprandial blood glucose responses (PPGRs) to meals based on individual factors, including dietary habits, physical activity, blood parameters, anthropometric data, and gut microbiome composition \cite{zeevi_personalized_2015}. After being trained on a cohort with 800 participants, GBR is validated using an independent cohort, achieving a Pearson correlation coefficient between predicted and measured PPGRs $R = 0.70$. A similar machine learning method has been used for other cohorts, such as Food \& You \cite{salathe2024personalized}.

Rein et al. extend this personalized approach to a clinical setting, focusing on a randomized dietary intervention pilot trial of 23 individuals with type 2 diabetes mellitus~\cite{rein2022effects}.  Based on the well-trained GBR, a personalized postprandial targeting diet is designed for each individual to minimize the individual’s PPGR.
Using a leave-one-out approach, the well-trained GBR assigns rankings to each participant's meals during the profiling week, where 4–6 distinct isocaloric options represent each meal type.

Neumann et al. predict the future blood glucose levels in type-1 diabetes patients during and after various types of physical activities in real-world conditions \cite{neumann2024data}. The study employs several machine learning and deep learning regression models, including XGBoost, Random Forest, LSTM, CNN-LSTM, and Dual-encoder models with an attention layer. The models use multiple data types, including continuous glucose monitoring data, insulin pump data, carbohydrate intake, exercise details (like intensity and duration), and physical activity-related information (e.g., number of steps and heart rate). The output is the predicted blood glucose level at future times, specifically at 10, 20, and 30 minutes after the inputs are recorded. Among many employed models, LSTM is the best-performing model for most patients. 
 
Although several machine-learning methods have been proposed to predict short-term postprandial responses of only a few metabolite biomarkers, less is explored over the important long-term responses of a wider range of health-related metabolites following dietary interventions. Wang et al. introduced a deep learning model, McMLP (Metabolic response predictor using coupled Multilayer Perceptrons), to fill this gap. McMLP consists of two coupled MLPs \cite{wang2025predicting}. The first MLP forecasts endpoint (i.e., after dietary interventions) microbial compositions from baseline (i.e., before dietary interventions) microbial and metabolomic profiles, and dietary intervention strategy. The second MLP uses these predicted endpoint microbial compositions, baseline metabolomic profiles as well as dietary intervention strategies to forecast endpoint metabolomic profiles. When McMLP is benchmarked with existing methods on synthetic data and six real data, it consistently yields a much better performance of predicting metabolic response than previous methods like random forest and GBR.

Managing blood glucose remains a major challenge for individuals with diabetes, as traditional methods often lack the adaptability required for personalized care. Recently, RL has emerged as a promising approach for optimizing insulin delivery in Type 1 Diabetes Mellitus (T1DM)~\cite{denes2024reinforcement}. The proposed system integrates continuous glucose monitoring (CGM) noise and random carbohydrate intake to emulate real-world variability. A closed-loop simulator was developed using the Identifiable Virtual Patient (IVP) model, to serve as the RL environment. A Proximal Policy Optimization (PPO) agent was trained within this framework, achieving an average Time in Range (TIR) of 73\%—demonstrating superior glucose control compared to conventional strategies. This study introduces a personalized, RL-driven insulin therapy solution that adapts dynamically to individual physiology and lifestyle. By leveraging PPO in a realistic simulation environment, it offers a scalable foundation for closed-loop artificial pancreas systems, with significant potential to enhance long-term glycemic outcomes in clinical practice.

Despite significant advancements in metabolic modeling and the integration of machine learning techniques for predicting metabolomic profiles, several open questions remain that could drive future research. One such question is to explore whether integrating multi-omics data (combining metagenomic, transcriptomic, and proteomic data) could further refine these predictions.

\subsection{Clinical Microbiology}
The earliest applications of AI in microbiology can be traced back to the 1970s when MYCIN was developed at Stanford University. MYCIN was an expert system designed to diagnose bacterial infections and recommend appropriate antibiotics. It used a rule-based approach, drawing on a knowledge base of expert-encoded rules to make decisions about infectious diseases, particularly blood infections. MYCIN was notable for demonstrating that AI could assist in clinical decision-making, setting the stage for later developments in AI for microbiology and medicine. AI pioneer Allen Newell referred to MYCIN as ``the granddaddy of expert systems'', stating it was ``the one that launched the field.'' Nowadays, various AI techniques have been applied in clinical microbiology. Here we briefly discuss those applications.  

\phantomsection
\subsubsection{Microorganism detection, identification and quantification}
AI techniques, especially supervised machine learning algorithms, are widely used to detect, identify, or quantify microorganisms using various types of data from cultured bacteria~\cite{peiffer2020machine}. Here we briefly discuss how AI techniques are applied across four different data types. (1) Microscopic Images: Deep learning models, particularly CNNs, have been highly effective in analyzing microscopic images of bacterial colonies~\cite{ramesh2024biointel,hallstrom2023label}. By training on labeled images, these models can classify bacterial species based on their shapes, sizes, arrangements, and staining characteristics (e.g., Gram staining). This approach aids in automating bacterial identification in clinical labs and research, improving the speed and accuracy of microbial diagnostics. (2) Spectroscopy Data: Supervised machine learning algorithms are also employed to analyze spectroscopy data, such as mass spectrometry or Raman spectroscopy, to identify microorganisms~\cite{wang2022identification,rahman2024machine}. For instance, MALDI-TOF (Matrix-Assisted Laser Desorption/Ionization Time-of-Flight) mass spectrometry generates unique protein ``fingerprints'' for bacterial species~\cite{clark2013matrix}. Machine learning models trained on these spectra can quickly and accurately classify species based on their spectral profiles. Raman spectroscopy, which provides molecular fingerprints of samples, also benefits from machine learning algorithms to classify bacterial species or detect specific metabolic or pathogenic profiles. (3) Volatile Organic Compounds (VOCs): Many bacteria emit VOCs as metabolic byproducts, which can serve as unique biomarkers for microbial identification~\cite{fend2006prospects}. Gas chromatography-mass spectrometry (GC-MS) or electronic noses (e-noses) are often used to capture these VOCs. Machine learning models trained on VOC patterns can distinguish bacterial species based on their unique VOC profiles. This approach has potential in medical diagnostics, food safety, and environmental monitoring. (4) Fecal microbial load. 
The microbiota in individual habitats differ in both relative composition and absolute abundance. While sequencing approaches determine the relative abundances of taxa and genes, they do not provide information on their absolute abundances. Recently, machine learning models, e.g., the eXtreme Gradient Boosting (XGBoost) regression model~\cite{nishijima2025fecal} and the Random Forest model~\cite{wirbel2025accurate} have been applied to predict fecal microbial loads solely from relative abundance data. It has been shown that fecal microbial load is the major determinant of gut microbiome variation and is associated with numerous host factors, including age, diet, and medication~\cite{nishijima2025fecal}.

Machine learning algorithms in these applications often require substantial labeled data for training, so accurate labeling and quality data collection are crucial. As these models learn to detect subtle differences in physical, chemical, and visual features, they contribute significantly to rapid, non-invasive, and automated bacterial identification, offering promising alternatives to traditional microbiological techniques.

\phantomsection
\subsubsection{Antimicrobial susceptibility evaluation}
The accurate and rapid determination of antimicrobial susceptibility is crucial for guiding effective therapy and combating the rise of multidrug-resistant pathogens. AI, particularly machine learning, has emerged as a powerful tool to accelerate and improve antimicrobial susceptibility testing (AST) by integrating genotypic, transcriptomic, and phenotypic data.

Early applications of AI focused on predicting phenotypic resistance directly from complex genotypic patterns. For example, MLP with one hidden layer trained on protease mutation profiles achieved high correlation ($R^2>0.88$) in predicting lopinavir resistance in human immunodeficiency virus
type 1 (HIV-1), revealing previously unrecognized resistance-associated mutations~\cite{wang2003enhanced}.

Machine learning has since been extended to bacterial pathogens. In \emph{Escherichia coli}, linear regression and feature-selection models trained on whole-genome sequencing (WGS) data accurately predict ciprofloxacin minimum inhibitory concentrations (MICs), with 93\% of predictions falling within a four-fold dilution range of the true value when using only four key mutation predictors~\cite{pataki2020understanding}. Similarly, in \emph{Pseudomonas aeruginosa}, multimodal classifiers combining gene presence/absence, sequence variation, and transcriptomic profiles achieved very high predictive values (>0.9) for resistance to four commonly administered antimicrobial drugs, demonstrating that gene expression data can further enhance diagnostic accuracy~\cite{khaledi2020predicting}.

A major breakthrough has been the fusion of genotypic and phenotypic information in a single rapid assay. GoPhAST-R simultaneously detects key resistance genes and early antibiotic-induced transcriptional changes, then applies a Random Forest classifier to classify susceptibility with 94–99\% accuracy~\cite{bhattacharyya2019simultaneous}. Performed directly from positive blood cultures using a hybridization-based multiplexed RNA detection platform, GoPhAST-R delivers phenotypic-equivalent AST results 24–36 hours faster than conventional culture-based methods, with total assay time under 4 hours.

Machine learning has also proven valuable in refining susceptibility breakpoints. For example, classification and regression tree (CART) analysis of pyrazinamide MIC in \emph{Mycobacterium tuberculosis} identified a clinical breakpoint of >50 mg/L predictive of treatment failure, aligning with independent hollow-fiber pharmacodynamic simulations and providing an evidence-based threshold for rapid molecular assays~\cite{gumbo2014pyrazinamide}.

Together, these studies illustrate a clear progression: from purely genotypic prediction of resistance phenotypes, to multimodal integration of genomic and expression data, and ultimately to rapid hybrid genotypic–phenotypic platforms that dramatically shorten time-to-result while maintaining or exceeding the accuracy of gold-standard methods. These AI-enabled approaches promise to transform routine microbiology diagnostics, enabling earlier targeted therapy and improved stewardship of existing antibiotics.

\phantomsection
\subsubsection{Disease diagnosis, classification, and clinical outcome prediction} 
AI can assist in examining novel and intricate data that clinical environments have not fully utilized for diagnostic aims. For instance, for certain diseases involving infections, microbes can generate some VOCs in clinical samples. Hence, we can utilize machine learning to evaluate the odors of those clinical samples to diagnose urinary tract infections~\cite{kodogiannis2008artificial}, active tuberculosis~\cite{mohamed2017qualitative}, pneumonia~\cite{he_exhaled_2024}, and acute exacerbation of chronic obstructive pulmonary disease~\cite{van2016diagnosing}.  For many other diseases associated with disrupted microbiomes, VOCs in clinical samples might not be helpful for disease diagnosis. In this case, we can leverage the microbiome data itself. Indeed, numerous studies have shown microbiome dysbiosis is associated with human diseases~\cite{lynch2016human,cryan2019microbiota}. Those diseases include GI disorders, i.e., \textit{Clostridioides difficile} infection~\cite{schubert2014microbiome}, inflammatory bowel disease~\cite{morgan2012dysfunction}, and irritable bowel syndrome~\cite{enck_irritable_2016}, and other non-GI disorders, for example, autism~\cite{kang2013reduced}, obesity~\cite{liu2015treatment}, multiple sclerosis~\cite{jangi2016alterations}, hepatic encephalopathy~\cite{kindt2018gut}, and Parkinson’s disease~\cite{scheperjans2015gut}. Applying supervised classification analysis to the human microbiome data can help us build classifiers that can accurately classify individuals' disease status, which could assist physicians in designing treatment plans~\cite{wu2021towards,han2025techniques}.

\phantomsection
\textbf{Classical machine learning classifiers.} 
Classical ML methods (e.g., Random Forest, XGBoost, Elastic Net, and SVM) have been systematically compared in the classification analysis of human microbiome data~\cite{wang_comparative_2020}. It was found that, overall, the XGBoost, Random Forest, and Elastic Net display comparable performance~\cite{wang_comparative_2020}. In case the training data contains microbiome data (features) collected before the disease diagnosis (labels), the well-trained classifiers can act as predictors, which have an even bigger clinical impact in terms of early diagnosis. For example, predicting asthma development at year three from the microbiome and other omics and clinical data collected at and before year one~\cite{wang2023benchmarking}.

\textbf{Phylogenetic tree-based deep learning methods.} 
Classical ML classifiers just treat microbiome data (more specifically, the taxonomic profiles) as regular tabular data, represented as a matrix with rows representing different samples or subjects and columns representing features (i.e., microbial species’ relative abundances). In fact, unlike many other omics, microbial features are endowed with a hierarchical structure provided by the phylogenetic tree defining the evolutionary relationships between those microorganisms. We can exploit the phylogenetic structure and leverage the CNN architecture to deal with species abundance data.  With this very simple idea, several deep learning methods (e.g., Ph-CNN~\cite{fioravanti_phylogenetic_2018}, PopPhy-CNN~\cite{reiman2020popphy}, taxoNN~\cite{sharma2020taxonn}, MDeep~\cite{wang_novel_2021} and PM-CNN~\cite{wang2024pm}) have been developed. Each method exploits the phylogenetic tree in a slightly different way. 

Ph-CNN takes the taxa abundances table and the taxa distance matrix as the input, and outputs the class of each sample~\cite{fioravanti_phylogenetic_2018}. Here, the distance between two taxa is defined as their patristic distance, i.e., the sum of the lengths of all branches connecting the two taxa on the phylogenetic tree. The patristic distance is used together with multi-dimensional scaling to embed the phylogenetic tree in an Euclidean space. Each taxon is represented as a point in Euclidean space preserving the tree distance as well as possible. Since the data is endowed with an intrinsic concept of neighborhood in the input space, Ph-CNN can then use CNN to perform classification. PopPhy-CNN represents the phylogenetic tree and species abundances in a matrix format, and then directly applies CNN to perform classification~\cite{reiman2020popphy}. taxoNN incorporates a stratified approach to group OTUs into phylum clusters and then applies CNNs to train within each cluster individually~\cite{sharma2020taxonn}. Further, through an ensemble learning approach, features obtained from each cluster were concatenated to improve prediction accuracy. Note that with each phylum cluster, the authors proposed two ways (either based on distance to the cluster center or based on taxa correlations) to order and place correlated taxa together to generate matrix or image-like inputs amenable for CNN.  MDeep directly incorporates the taxonomic levels of the phylogenetic tree into the CNN architecture~\cite{wang_novel_2021}. OTUs on the species level are clustered based on the evolutionary model. This clustering step makes convolutional operation capture OTUs highly correlated in the phylogenetic tree. The number of hidden nodes decreases as the convolutional layer moves forward, reflecting the taxonomic grouping. PM-CNN (Phylogenetic Multi-path Convolutional Neural Network) is a phylogeny-based neural network model designed for multi-status classification and disease detection using microbiome data~\cite{wang2024pm}. It organizes microbes according to their phylogenetic relationships and extracts features through a multi-path CNN, a deep learning architecture that processes data via parallel paths. Unlike traditional CNNs, which rely on a single path to extract features layer by layer, potentially leading to information loss due to limited feature representation in complex tasks, the multi-path CNN processes input data through multiple independent paths. This enables feature extraction from diverse angles and scales, enhancing the model's expressiveness and robustness for complex tasks. In PM-CNN, the the microbial richness features are embedded with phylogeny by multi-layer hierarchical clustering, which are then fed into a multi-path CNN for disease classification. It has been shown that PM-CNN outperforms existing machine learning models in microbiome-based disease classification.

\textbf{Other deep learning methods.} 
Besides the above deep learning methods that exploit the phylogenetic structure for microbiome data classification, some other deep learning methods (e.g., DeepMicro~\cite{liang_deepmicrobes_2020}, GDmicro~\cite{liao2023gdmicro},  and a transformer-based microbial ``language'' model~\cite{pope2023learning}) have been developed. Those methods do not leverage the phylogenetic structure of microbiome data. 

DeepMicro incorporated various autoencoders (including SAE, DAE, VAE, and CAE) to learn a low-dimensional embedding for the input microbial compositional feature, and then employed MLP to classify disease status with the learned latent features~\cite{oh2020deepmicro}. GDmicro is a GCN-based method for microbiome feature learning and disease classification~\cite{liao2023gdmicro}.  
GDmicro formulates the disease classification problem as a semi-supervised learning task, which uses both labeled and unlabeled data for feature learning (\cite{van2020survey}). To overcome the domain discrepancy problem (i.e., data from different studies have many differences due to confounding factors, such as region, ethnicity, and diet, which all shape the gut microbiome), GDmicro applies a deep adaptation network~\cite{long2015learning} to learn transferable latent features from the microbial compositional matrix across different domains/studies with or without disease status labels. Then, GDmicro constructs a similarity graph, where each node represents a host whose label can be either healthy, diseased, or unlabeled, and edges represent the similarity between two hosts’ learned latent features. GDmicro then employs GCN to take this microbiome similarity graph as input and incorporate both the structural and node abundance features for disease status classification. Note that this is a very classical application of GCN to solve the semi-supervised node classification problem on graphs, where some nodes have no labels.  

Recently, a transformer-based microbial ``language'' model (MLM) was developed~\cite{pope2023learning}. This MLM was trained in a self-supervised fashion to capture the interactions among different microbial species and the common compositional patterns in microbial communities. The trained MLM can generate robust, context-sensitive representations of microbiome samples to enhance predictive modeling. Note that in this transformer-based MLM, taxa present in each microbiome sample were ranked in decreasing order of abundance to create an ordered list of taxa so that the inputs are analogous to texts. The transformer model then processes these inputs through multiple encoder layers, producing a hidden representation for each taxon. The output of the model includes both sample-level embeddings for classification tasks and context-sensitive embeddings for individual taxon, enabling a nuanced understanding of microbial interactions. By pre-training the transformer using self-supervised learning on large, unlabeled datasets and fine-tuning on specific labeled tasks, this approach leads to improved performance for multiple prediction tasks including predicting IBD and diet patterns. Multistage Fusion Tabular (MSFT) Transformer~\cite{wang2025msft} is another transformer-based model designed for disease prediction by effectively integrating diverse, high-dimensional tabular data derived from metagenomic data. Its multistage fusion strategy comprises three key modules: an early-stage fusion-aware feature extraction module that enhances the quality of extracted input information, a mid-stage alignment-enhanced fusion module that ensures the retention of critical information during cross-modal learning, and a late-stage integrated feature decision layer that consolidates cross-modal information for robust predictions.

Despite the development of various methods, a systematical comparison of those deep learning methods and classical machine learning methods on benchmarking datasets is lacking. Since some of those deep learning methods incorporate domain knowledge (i.e., information on the phylogenetic tree, or unlabeled samples), it would be necessary to do that for classical ML methods too, for a fair comparison. In addition, while recent deep learning models often outperform traditional machine learning approaches, these advantages typically depend on the availability of large, high-quality labeled datasets and substantial computational resources. In contrast, traditional methods (e.g., random forest, support vector machines) can offer more robust performance in small-sample or low-resource scenarios, where deep learning models are prone to overfitting and require costly training. Thus, model selection should consider the specific data characteristics: deep learning excels with abundant data and computational power, whereas classical machine learning approaches remain valuable for smaller or noisier datasets.

{\bf Integration of various feature types.} 
Note that 16S rRNA gene sequencing can only provide taxonomic profiles (in terms of microbial compositions) and cannot directly profile microbial genes/functions. Shotgun metagenome sequencing can provide comprehensive data on both taxonomic and functional profiles. It is quite natural to investigate if combining both taxonomic and functional features will enhance classification performance. MDL4Microbiome is such a deep learning method. It employs MLP and combines three different feature types, i.e., taxonomic profiles, genome-level relative abundance, and metabolic functional characteristics, to enhance classification accuracy~\cite{lee_multimodal_2022}.

Quite often, we have multi-omics data and clinical data. It would be more insightful to integrate those different data types for better disease status classification or prediction~\cite{monshizadeh2024incorporating}. A straight approach would be to concatenate all datasets into a single view, which is then used as the input to a supervised learning model of choice. A more advanced approach is MOGONET, which jointly explores omics-specific learning using GCNs and cross-omics correlation learning for effective multi-omics data classification~\cite{wang_mogonet_2021}.

Recently, in a childhood asthma prediction project, 18 methods were evaluated using standard performance metrics for each of the 63 omics combinations of six omics data (including GWAS, miRNA, mRNA, microbiome, metabolome, DNA methylation) collected in The Vitamin D Antenatal Asthma Reduction Trial cohort~\cite{wang2023benchmarking}. It turns out that, surprisingly, Logistic Regression, MLP, and MOGONET display superior performance than other methods. Overall, the combination of transcriptional, genomic, and microbiome data achieves the best prediction for childhood asthma prediction. In addition, including the clinical data (such as the father and mother’s asthma status, race, as well as vitamin D level in the prediction model) can further improve the prediction performance for some but not all the omics combinations. Results from this study imply that deep learning classifiers do not always outperform traditional classifiers.

So far, the integration of various data types discussed above is often referred to as early fusion. It begins by transforming all datasets into a single representation, which is then used as the input to a supervised learning model of choice.  There is another approach called late fusion, which works by developing first-level models from individual data types and then combining the predictions by training a second-level model as the final predictor. Recently, encompassing early and late fusions, cooperative learning combines the usual squared error loss of predictions with an agreement penalty term to encourage the predictions from different data views to align~\cite{ding2022cooperative}. It would be interesting to explore this idea of cooperative learning in disease classification using multi-omics data~\cite{meqdad2023classification,ferjani2020cooperative} (including microbiome data).

\subsection{Prevention \& Therapeutics}
\phantomsection
\subsubsection{Vaccine design}
Vaccines work by stimulating the immune system to produce antibodies, offering protection against future infections. Traditional vaccine development, known as vaccinology, involves isolating a pathogen, identifying its antigenic components, and testing them for immune response. Reverse vaccinology (RV), a more modern and computational approach, begins by analyzing the pathogen’s genome to identify potential antigenic proteins, which are then synthesized and evaluated as vaccine candidates. RV accelerates vaccine discovery and can reveal novel targets that traditional methods might overlook~\cite{pizza2000identification,dalsass2019comparison}.

Current RV approaches can be classified into two categories: (1) rule-based filtering methods, e.g., NERVE~\cite{vivona2006nerve} and Vaxign~\cite{he2010vaxign}; and (2) Machine learning-based methods, e.g., VaxiJen~\cite{doytchinova2007vaxijen}, ANTIGENpro~\cite{magnan2010high}, Antigenic~\cite{rahman2019antigenic}, and Vaxign-ML~\cite{ong2020vaxign,ong2021vaxign2}. The rule-based filtering method narrows down potential vaccine candidates from the large number of antigenic proteins identified through genome analysis. This process involves applying predefined biological rules or criteria (e.g., protein localization, the absence of similarity to host proteins to reduce the risk of autoimmune responses, immunogenicity potential, etc.). These rules help prioritize proteins most likely to elicit a protective immune response, speeding up vaccine candidate identification. Note that all these currently available rule-based filtering methods use only biological features as the data input. Machine learning-based RV methods predict potential vaccine candidates by training classifiers on known antigenic proteins and non-antigenic proteins. These machine learning methods can analyze physicochemical or biological features of the input proteins, and then classify new proteins based on the learned patterns. These machine learning methods can identify vaccine candidates with higher accuracy and efficiency compared to traditional methods, leveraging vast datasets and complex patterns that may not be evident through rule-based filtering alone. For example, Vaxign-ML, the successor to Vaxign, utilized XGBoost as the classifier and emerged as the top-performing Machine learning-based RV methods~\cite{ong2020vaxign,ong2021vaxign2}. 

Recently, deep learning techniques have also been developed for RV. For example, Vaxi-DL is a web-based deep learning software that evaluates the potential of protein sequences to serve as vaccine target antigens~\cite{rawal2022vaxi}. Vaxi-DL consists of four different deep learning pathogen models trained to predict target antigens in bacteria, protozoa, fungi, and viruses, respectively.  All the four pathogen models are based on MLPs. For each pathogen model, a particular training dataset consisting of antigenic (positive samples) and non-antigenic (negative samples) sequences was derived from known vaccine candidates and the Protegen database. Vaxign-DL is another deep learning-based method to predict viable vaccine candidates from protein sequences~\cite{zhang2023vaxign}. Vaxign-DL is also based on MLP. It has been shown that Vaxign-DL achieved comparable results with Vaxign-ML in most cases, and outperformed Vaxi-DL in the prediction of bacterial protective antigens. 

In the future, it would be interesting to test if other deep learning models (e.g., 1D CNN, RNN, and its variants, or Transformer) can also be used to predict target antigens.

\phantomsection
\subsubsection{Probiotic mining}
The discovery and experimental validation of probiotics demand significant time and effort. Developing efficient screening methods for identifying probiotics is therefore of great importance. Recent advances in sequencing technology have produced vast amounts of genomic data, allowing us to design machine learning-based computational approaches for probiotic mining. For example, Sun et al. developed iProbiotics, which utilizes $k$-mer frequencies to characterize complete bacterial genomes and employs the support vector machine for probiotic identification~\cite{sun_iprobiotics_2022}. iProbiotics conducted a $k$-mer compositional analysis (with $k$ ranging from 2 to 8) on a comprehensive probiotic genome dataset, which was built using the PROBIO database and literature reviews. This analysis revealed significant diversity in oligonucleotide composition among strain genomes, showing that probiotic genomes exhibit more probiotic-related features compared to non-probiotic genomes. A total of 87,376 $k$-mers were further refined using an incremental feature selection method, with iProbiotics achieving peak accuracy using 184 core features. This study demonstrated that the probiotic role is not determined by a single gene but rather by a composition of $k$-mer genomic elements. 

Although iProbiotics has been validated using complete bacterial genomes, its effectiveness on draft genomes derived from metagenomes remains uncertain. Additionally, while the $k$-mer frequency model has been applied in various bioinformatics tasks, it primarily captures the occurrence frequencies of oligonucleotides and may not fully represent sequence function. Recent advancements in NLP have introduced novel methods for representing biological sequences. In these models, oligonucleotides or oligo-amino acids are treated as 'words,' and DNA or protein sequences as 'sentences.' By using unsupervised pretraining on large datasets, each word is mapped to a context-based feature vector, potentially offering more informative representations than $k$-mer frequencies. Building on this concept, Wu et al. developed metaProbiotics, a method designed to mine probiotics from metagenomic binning data~\cite{wu2024metaprobiotics}. It represents DNA sequences in metagenomic bins using word vectors and employs random forests to identify probiotics from the metagenomic binned data.

Technically speaking, both iProbiotics and metaProbiotics are not based on deep learning techniques. In particular, the classification analysis still relies on traditional machine learning methods, e.g., SVM and RF. We expect that soon more deep learning-based methods will be developed to solve this very important task.

\phantomsection
\subsubsection{Antibiotic discovery} 
Compared with probiotic discovery, deep learning has been extensively used in antibiotic discovery. This thanks to the success of GCNs, which have been repeatedly shown to have robust capacities for modeling graph data such as small molecules. In particular, message-passing neural networks (or MPNNs) are a group of GCN variants that can learn and aggregate local information of molecules through iterative message-passing iterations~\cite{gilmer2017neural}. MPNNs have exhibited advancements in molecular modeling and property prediction.

The original MPNN operates on undirected graphs. It is trivial to extend MPNN to directed multigraphs. This yields Directed MPNN (or D-MPNN), which translates the graph representation of a molecule into a continuous vector via a directed bond-based message passing approach~\cite{yang2019analyzing}. This builds a molecular representation by iteratively aggregating the features of individual atoms and bonds. The model operates by passing ``messages'' along bonds that encode information about neighboring atoms and bonds. By applying this message passing operation multiple times, the model constructs higher-level bond messages that contain information about larger chemical substructures. The highest-level bond messages are then combined into a single continuous vector representing the entire molecule.

D-MPNN has recently been utilized to discover novel antibiotics~\cite{stokes2020deep,wong2024discovery}. For each compound, RDKit generates a graph-based molecular representation from its Simplified Molecular-Input Line-Entry System (SMILES) string. Feature vectors for atoms and bonds are created using computable properties: atom features include atomic number, number of bonds, formal charge, chirality, number of bonded hydrogen atoms, hybridization, aromaticity, and atomic mass; bond features encompass bond type (single, double, triple, or aromatic), conjugation, ring membership, and stereochemistry. During message passing, each bond-associated message (a real number) is updated by summing messages from neighboring bonds, concatenating the current bond’s message with this sum, and applying a neural network layer with a nonlinear activation function. After a fixed number of message-passing steps, the molecule’s messages are aggregated to produce a final molecular representation. This representation is processed through a feed-forward neural network to predict the compound’s activity, such as antibiotic efficacy, cytotoxicity, or proton motive force-altering potential.

The Atom-Bond Transformer-based Message-Passing Neural Network (ABT-MPNN)~\cite{liu2023abt}, a novel variant of MPNN, also holds great potential for antibiotic discovery. ABT-MPNN integrates the self-attention mechanism of Transformers with the MPNN framework to enhance molecular representation and improve molecular property predictions. By incorporating tailored attention mechanisms during the message-passing and readout phases, ABT-MPNN provides an innovative end-to-end architecture that unifies molecular representations at the bond, atom, and molecule levels. Additionally, the model offers a visualization modality for attention at the atomic level, providing valuable insights into the molecular atoms or functional groups associated with desired biological properties. This feature makes ABT-MPNN a powerful tool for investigating the mechanisms of action of drugs, including but not limited to antibiotics.

While deep learning approaches effectively identify antibacterial compounds from existing libraries, their structural novelty remains limited. Recently, a deep learning-based framework was developed to design novel antibiotic compounds using either fragment-based or de novo approaches~\cite{krishnan2025generative}. In the fragment-based approach, D-MPNN screened over 45 million chemical fragments in silico, identifying those with predicted selective antibacterial activity against \emph{Neisseria gonorrhoeae} and \emph{Staphylococcus aureus}. These fragments were then expanded into complete molecules using two generative algorithms: a genetic algorithm based on Chemically Reasonable Mutations (CReM) and a VAE. CReM generates new molecules by adding, replacing, or deleting atoms and functional groups starting from a compound of interest. In the de novo approach, no fragment input is required; instead, CReM and VAE design molecules based on patterns learned during training. This framework enables generative deep learning-guided design of antibiotic candidates, facilitating the discovery of novel antibacterial compounds and efficient exploration of vast, uncharted chemical space.

Experimental determination of antimicrobial activity for novel compounds remains time-consuming and costly. While compound-centric deep learning models can accelerate candidate prioritization, they typically require large, custom-labeled training datasets. Recently, a lightweight, data-efficient strategy for antimicrobial discovery was introduced~\cite{olayo2025pre}. This approach builds on MolE (Molecular representation through redundancy-reduced Embedding), a self-supervised framework that learns task-agnostic molecular embeddings from unlabeled chemical structures. By leveraging Graph Isomorphism Networks (GINs) and adapting the non-contrastive Barlow-Twins objective to the molecular domain, MolE produces robust, generalizable representations. When integrated with experimentally validated compound–bacteria activity data, these embeddings enable a high-performing XGBoost-based model to predict the antimicrobial potential of any compound. This scalable, cost-effective pipeline significantly accelerates antibiotic discovery.

\phantomsection
\subsubsection{Antimicrobial peptides identification \& design}
Bacterial resistance to antibiotics is a growing concern. 
Antimicrobial peptides (AMPs) are a class of small peptides (containing fewer than 100 amino acids) that widely exist in nature~\cite{huan2020antimicrobial}. AMPs are an important part of the innate immune system of different organisms and hold the potential to tackle multidrug-resistant pathogens. In recent years, AMPs have been considered as a substitute for traditiona antibiotics due to their unique antibacterial mechanism.

AMP identifications through wet-lab experiments is very expensive and time consuming. Various machine learning methods have been developed to facilitate this process. Early attempts leveraged traditional classifiers, shallow neural networks, or antimicrobial scoring functions, using with well-selected features. 
For example, AntiBP~\cite{lata2007analysis} used Support Vector Machine, Quantitative Matrices and MLP with only one hidden layer to predict AMPs based on N-terminal or/and C-terminal residues. The feature selection was inspired by the observation that certain types of residues are preferred at the N-terminal (or C-terminal) regions of the AMPs. 
In another work~\cite{torrent2011connecting}, MLP with one hidden layer was used to identify AMPs based on their physicochemical properties summarized in eight features: isoelectric point (pI), peptide length, $\alpha$-helix, $\beta$-sheet and turn structure propensity, \emph{in vivo} and \emph{in vitro} aggregation propensity and hydrophobicity. 
iAMP-2L~\cite{xiao2013iamp} used a fuzzy K-nearest neighbor algorithm and the pseudo–amino acid composition of AMPs featured by five physicochemical properties of amino acids: (1) hydrophobicity; (2) pK1(C$^\alpha$-COOH), (3) pK2(NH3); (4) pI(\SI{25}{\celsius}); and (5) molecular weight. 
iAMPpred~\cite{meher2017predicting} used SVM with three compositional (amino acid composition-AAC, pseudo amino acid composition-PAAC and normalized amino acid composition-NAAC), three physicochemical (hydrophobicity, net-charge and isoelectric point) and three structural ($\alpha$-helix propensity, $\beta$-sheet propensity and turn propensity) features to predict AMPs. 
AmPEP~\cite{bhadra2018ampep} used Random Forest and the distribution patterns of amino acid properties (including hydrophobicity, normalized van der Waals volume, polarity, polarizability, charge, secondary structure, and solvent accessibility) to predict AMPs. 
An antimicrobial scoring function~\cite{pane2017antimicrobial} integrating key physicochemical determinants of AMPs, including net charge, average hydrophobicity, and sequence length, was developed to identify cationic AMPs (CAMPS). This scoring function is based on the empirical observation that the antimicrobial potency of CAMPs is linearly correlated to the product $C^m H^n L$, where $C$, $H$, and $L$ represent the net charge,  hydrophobicity, and length of the peptide, respectively. Exponents $m$ and $n$ define the relative contribution of the net charge and hydrophobicity to the antimicrobial potency. This scoring function can be used to detect CAMPs included inside the structure of larger proteins or precursors. Indeed, it has been recently applied to mine the human proteome to identify encrypted AMPs --- many encoded within proteins unrelated to immune function~\cite{torres2022mining}.

Note that those methods work on peptide sequences and are not directly applicable to microbial genomes or metagenomes. Macrel --- (Meta)genomic AMP classification and retrieval is the first end-to-end pipeline for the prospection of high-quality AMP candidates from (meta)genomes~\cite{santos2020macrel}. In particular, Macrel accepts as inputs paired-end or single-end reads in FastQ format, performs quality-based trimming (using NGLess~\cite{coelho2019ng}), assembles contigs (using MEGAHIT~\cite{li2016megahit}), and predicts smORFs (using a modified version of Prodigal~\cite{hyatt2010prodigal}). The resulting smORFs (10~100 amino acids) will be used for AMP classification. Marcel combines 6 local and 16 global features, and uses Random Forest for AMP identification. It has been demonstrated that Marcel is comparable to many existing methods with different trade-offs. In particular, Macrel achieves the highest precision and specificity at the cost of lower sensitivity. This property of Macrel (i.e., emphasizing on precision over sensitivity) has been leveraged recently to identify AMPs in the global microbiome (a vast dataset of 63,410 metagenomes and 87,920 prokaryotic genomes from environmental and host-associated habitats) and create the so-called AMPSphere~\cite{santos2024discovery}, a comprehensive catalog comprising 863,498 nonredundant peptides.

Those AMP prediction methods described above require quite a lot of domain knowledge and manual feature selection. This effort can be avoided or mitigated by using deep learning models that can automatically learn complex representations and features from raw data, reducing the need for manual feature engineering. For example, AMPScanner~\cite{veltri2018deep} is a deep neural network model with convolutional and recurrent layers that leverage primary sequence composition. By combining CNN and RNN, AMPScanner can extract more meaningful and robust features and avoid the burden of a priori feature construction. MLBP~\cite{tang2022identifying} is another similar hybrid deep learning model that integrated CNN and RNN to predict AMPs. Firstly, the amino acids were converted into natural numbers, and the sequences of all peptides were set to be fixed by using the zero-filled method. Then, an embedding layer was used to learn the embedding matrix of the representation of peptide sequences. The embedding matrix was fed into a CNN to extract the features from the peptide. Then, an RNN is used to analyze streams of the sequence by means of hidden units. Finally, a fully connected layer is applied to the final classification. 
AMPlify~\cite{li2022amplify} is an attention-enhanced deep learning model that integrates two complementary attention mechanisms (multi-head scaled dot-product attention (MHSDPA) and context attention (CA)) atop a bidirectional LSTM backbone. The bidirectional LSTM first encodes the input sequence recurrently, capturing rich positional and contextual dependencies. The MHSDPA layer then refines this representation by computing multiple weighted views of the sequence using distinct attention heads. Finally, the CA layer produces a single, informative summary vector through a learned weighted average, distilling key contextual insights from the preceding layer. Note that AMPScanner, MLBP, and AMPlify take the peptide sequence as the input. Recently, an end-to-end pipeline for AMP prediction from the human gut microbiome data using deep learning was developed~\cite{ma_identification_2022}. Instead of using any manually selected features, the authors combined five deep learning models, including (1) Two CNN + LSTM models; (2) Two CNN + Attention models; and (3) One BERT model to automatically extract features for AMP prediction. Because the prediction biases were independent of each other, the authors eventually tested the intersection of predictions from various combinations of those models. This can be considered as an ensemble approach. Their pipeline starts with the prediction of smORFs from 4,409 high-quality representative genomes of microbial organisms present in human microbiome samples. Those genomes were assembled in a previous study~\cite{pasolli2019extensive}. Then they predicted 20,426,401 putative AMPs from the non-redundant smORFs using their ensemble approach. To ensure the predicted AMPs are indeed expressed, they leveraged multiple metaproteome datasets to further identify 2,349 candidate AMPs. Finally, they performed correlation analysis between the abundances of those candidate AMPs and bacteria, using metagenomic datasets from 15 independent large cohorts. They hypothesized that those candidate AMPs with strong negative correlations with members of a microbiome are more likely to be functional. Although this study focused on the human gut microbiome, the developed pipeline can be easily adopted for AMP mining in other microbiomes.

Molecular de-extinction (MDE) is the resurrection of extinct biomolecules that are no longer encoded by living organisms. MDE is driven by the hypothesis that molecules conferring evolutionary advantages to extinct species may offer benefits in the modern global environment. Recently, MDE has emerged as a novel framework for AMP discovery by expanding the therapeutic search space through paleoproteome mining. Specifically, panCleave~\cite{maasch2023molecular}, a Random Forest-based model, was developed to perform proteome-wide cleavage site prediction and computational proteolysis (i.e., the in silico digestion of human proteins into peptide fragments) to prospect AMPs encrypted within extinct and extant human proteins. In vitro antimicrobial activity was confirmed for both modern and archaic protein-derived peptides identified by panCleave. Lead candidates demonstrated proteolytic resistance and variable membrane permeabilization. Recently, a deep learning method APEX has been developed to mine proteomes of extinct species~\cite{wan2024deep} and archaeal organisms~\cite{torres2025deep} for the discovery of AMPs. APEX utilized a hybrid of recurrent and attention neural networks to extract peptide sequence information, which was further processed by two fully connected neural networks to predict species-specific antimicrobial activities (i.e., a multi-output regression task) or general antimicrobial or not (i.e., a binary classification task), respectively. To improve the prediction performance, an ensemble learning approach was adopted by selecting the top eight APEX models (with different neural network architectures and training strategies). The guidelines of using machine learning to mine genomes and proteomes for AMP discovery have been summarized nicely in a tutorial~\cite{wan2025tutorial}.

Besides identifying natural AMPs, deep learning approaches have also been developed to design synthetic AMPs. Early attempts leveraged GAN and VAE, as well as their conditional variants cGAN and cVAE. For example, AMPGANv2 is based on a bidirectional conditional GAN~\cite{van2021ampgan}. It uses generator-discriminator dynamics to learn data-driven priors and control generation using conditioning variables~\cite{van2021ampgan}. The bidirectional component, implemented using a learned encoder to map data samples into the latent space of the generator, aids iterative manipulation of candidate peptides. These elements allow AMPGANv2 to generate candidates that are novel, diverse, and tailored for specific applications. Training of GANs was reported to face substantial technical obstacles, such as training instabilities and mode collapse. VAE-based AMP generations could be an alternative solution. For example, Peptide VAE is based on a VAE, where both encoder and decoder are single-layer LSTMs~\cite{dean2021pepvae}. The authors also proposed Conditional Latent (attribute) Space Sampling (CLaSS) for controlled sequence generation, aimed at controlling a set of binary (yes/no) attributes of interest, such as antimicrobial function and/or toxicity. HydrAMP is based on a conditional VAE to generate novel peptide sequences satisfying given antimicrobial activity conditions~\cite{szymczak_discovering_2023}. This method is suitable not only for the generation of AMPs de novo, but also for the generation starting off from a prototype sequence (either known AMPs or non-AMPs). Recently, a conditional denoising VAE (cdVAE) model was developed for AMP generation~\cite{zhao2025conditional}. For each AMP sample, 10 physicochemical properties—molecular weight, isoelectric point, GRAVY, aromaticity, instability index, disulfide bonds, molecular volume, and secondary structure fractions ($\alpha$-helix, $\beta$-sheet, and random coil)—are computed and normalized to form a property vector. The AMP sequence is one-hot encoded and passed through an embedding layer to generate an embedding vector. Gaussian noise from a normal distribution is added to this embedding vector to create noised samples. These noised samples, along with the property vector, are processed through the cdVAE’s encoder and decoder, producing a reconstructed sample that achieves a denoising effect. Specifically, the encoder maps the AMP input data distribution to a standard normal distribution, while the decoder transforms samples from this distribution into AMP sequences with desired physicochemical properties. Notably, the property vector guides the VAE training process, enabling the decoder to map the standard normal distribution to a conditional distribution of AMPs with tailored physicochemical characteristics.

EvoGradient~\cite{wang2025explainable} (Evolution guided by Gradient) is an interesting attempt to virtually modify peptide sequences to produce more potent AMPs, akin to in silicon directed evolution. This was achieved through an iterative process that combines gradient descent with projection to find a locally optimal sequence near the original peptide. In each step, a scaled gradient is subtracted from the input vector, followed by a projection back to one-hot format to check for any amino acid mutations. If a mutation is found, the projected result is used for the next iteration; otherwise, projection is skipped. This amino acid editing is performed iteratively until half of the original peptide’s amino acids are altered or a local optimum is reached. deepAMP~\cite{li2024foundation}, which takes peptides with low antimicrobial
activity as inputs and output analogs with high antimicrobial activity and broad-spectrum resistance. It achieves this through a pre-training and multi-stage fine-tuning strategy, coupled with a sequence degradation approach for data augmentation. Promising results demonstrate deepAMP's ability to accelerate the discovery of effective AMPs against drug-resistant bacteria, including \emph{Staphylococcus aureus}, \emph{Klebsiella pneumoniae}, and \emph{Pseudomonas aeruginosa}. AMPainter~\cite{dong2025painting} is a deep reinforcement learning framework that unifies optimization and generation of AMPs in a single pipeline. It is applied to three peptide classes: known AMPs, signal peptides (SPs), and random sequences. AMPainter outperforms ten benchmark models in enhancing predicted antimicrobial potency and sequence diversity when optimizing known AMPs. Given a set of input peptide sequences, AMPainter iteratively evolves them into improved AMPs through a three-step process: (1) a policy network identifies mutation sites, (2) a fine-tuned language model proposes residue replacements, and (3) HyperAMP—a hypergraph neural network pre-trained on known AMPs and their activity labels—evaluates antimicrobial potential. The predicted activity scores serve as rewards to update the policy network via reinforcement learning, enabling continuous improvement across iterations.

TransSAFP~\cite{liu2025novo} is a deep learning model that predicts the functional activity of self-assembling functional peptides (SAFPs) with minimal experimental annotation. This enables de novo design of AMPs. TransSAFP comprises two modules: a pretraining module and a transfer learning module. The pretraining architecture is a multi-head self-attention transformer with a cross-attention layer, while the downstream transfer learning module appends a self-attention block to the pretrained latent representation. The resulting model designed a lead self-assembling peptide that demonstrates excellent in vivo therapeutic efficacy against intestinal bacterial infections in mice, with superior biofilm eradication and no induction of acquired drug resistance. Mechanistic studies reveal that the peptide self-assembles on bacterial membranes to form nanofibrous structures, effectively killing multidrug-resistant pathogens. TransSAFP is promising for the customized design of functional peptide materials.

\phantomsection
\subsubsection{Phage therapy}
As the most abundant organisms in the biosphere, bacteriophages (a.k.a. phages) are viruses that specifically target bacteria. They play a significant role in microbial ecology by influencing bacterial populations, gene transfer, and nutrient cycles. Moreover, they can be an alternative to antibiotics and hold the potential therapeutic ability for bacterial infections~\cite{schooley2017development,pirnay2024personalized,green2023retrospective,chen2023virbot,de2025exploring}.

\textbf{Phage identification.} Many computational tools have been developed to identify bacteriophage sequences in metagenomic datasets~\cite{ho2023gauge}.  They can be roughly grouped into two classes: (1) alignment-based (or database-based) methods, e.g., MetaPhinder~\cite{jurtz2016metaphinder}, VIBRANT~\cite{kieft2020vibrant}, and VirSorter2~\cite{guo2021virsorter2}; (2) alignment-free methods, e.g., VirFinder~\cite{ren2017virfinder}, PPR-meta~\cite{fang2019ppr}, Seeker~\cite{auslander2020seeker}, DeepVirFinder~\cite{ren2020identifying}, and PhaMer~\cite{shang2022accurate}. Alignment-based methods typically use a large number of sequences of references and utilize DNA or protein sequence similarity as the main feature to distinguish phages from other sequences. Their limitations are evident. Firstly, bacterial contigs may align with multiple phage genomes, potentially resulting in false-positive phage predictions. Secondly, novel or highly diverged phages may not have significant alignments with the selected phage protein families, which can lead to low sensitivity in identifying new phages. Alignment-free methods can overcome those limitations via machine learning or deep learning techniques. Those methods learn the features of the sequence data and are mainly classification models with training data consisting of both phages and bacteria.  Some classification models use manually extracted sequence features such as k-mers, while others use deep learning techniques to automatically learn features. For example, VirFinder uses $k$-mers to train a logistic regression model for phage identification. Seeker (or DeepVirFinder) uses one-hot encoding to represent the sequence data and trains an LSTM (or CNN) to identify phages, respectively. PhaMer leverages the start-of-the-art language model, the Transformer, to conduct contextual embedding for phage contigs. It feeds both the protein composition and protein positions from each contig into the Transformer, which learns the protein organization and associations to predict the label for test contigs. It has been shown that PhaMer outperforms VirSorter, Seeker, VirFinder, DeepVirfinder, and PPR-meta. geNomad is a hybrid framework that combines the strengths of alignment-free and alignment-based models for concurrent identification and annotation of both plasmids and viruses in sequencing data~\cite{camargo2023identification}. To achieve that, geNomad processes user-provided nucleotide sequences via two distinct branches. In the sequence branch (``alignment-free''), the inputs are one-hot encoded and passed through an IGLOO neural network, which evaluates them by identifying non-local sequence motifs. In the marker branch (``alignment-based''), the proteins encoded by the input sequences are annotated with markers specific to chromosomes, plasmids, or viruses. Here, the key idea behind the IGLOO neural network is to leverage the relationships between ``non-local patches'' sliced from feature maps generated by successive convolutions to effectively represent long sequences, allowing it to handle both short and long sequences efficiently, unlike traditional RNNs which struggle with very long sequences~\cite{sourkov2018igloo}.

Recently, a hybrid method called INHERIT was developed. INHERIT (IdentificatioN of bacteriopHagEs using deep RepresentatIon model with pre-Training) naturally ‘inherits’ the characteristics from both alignment-based and alignment-free methods~\cite{bai2022identification}. In particular, INHERIT uses pre-training as an alternative way of acquiring knowledge representations from existing databases, and then uses a BERT-style deep learning framework to retain the advantage of alignment-free methods.  The independent pre-training strategy can effectively deal with the data imbalance issue of bacteria and phages, helping the deep learning framework make more accurate predictions for both bacteria and phages. The deep learning framework in INHERIT is based on a novel DNA sequence language model: DNABERT~\cite{ji_dnabert_2021}, a pre-trained bidirectional encoder representation model, which can capture global and transferrable understanding of genomic DNA sequences based on up and downstream nucleotide contexts. It has been demonstrated that INHERIT outperforms four existing state-of-the-art approaches: VIBRANT, VirSorter2, Seeker, and DeepVirFinder. It would be interesting to compare the performance of INHERIT and PhaMer. Also, since DNABERT is per-trained on human genomes, it would be interesting to see if INHERIT can be further improved when using a BERT model pre-trained on prokaryotic genomes.

\textbf{Phage lifestyle prediction.} Besides phage identification, machine learning techniques can also be used to predict the phage lifestyle (virulent or temperate), which is crucial to enhance our understanding of the phage-host interactions. For example, PHACTS used an RF classifier on protein similarities to classify phage lifestyles~\cite{mcnair2012phacts}. BACPHLIP also used an RF classifier on a set of lysogeny-associated protein domains to classify phage lifestyles~\cite{hockenberry2021bacphlip}. Those two methods do not work well for metagenomic data. By contrast, DeePhage can directly classify the lifestyle for contigs assembled from metagenomic data~\cite{wu_deephage_2021}. DeePhage uses one-hot encoding to represent DNA sequences and trains a CNN to obtain valuable local features. PhaTYP further improved the accuracy of phage lifestyle prediction on short contigs by adopting BERT to learn the protein composition and associations from phage genomes~\cite{shang2023phatyp}. In particular, PhaTYP solved two tasks: a self-supervised learning task and a fine-tuning task. In the first task, PhaTYP applies self-supervised learning to pre-train BERT to learn protein association features from all the phage genomes, regardless of the available lifestyle annotations. In the second task, PhaTYP fine-tunes BERT on phages with known lifestyle annotations for classification. It has been shown that PhaTYP outperforms DeePhage and three other machine learning methods PHACTS (based on RF), BACPHLIP (based on RF), and PhagePred (based on Markov model). DeePhafier is another deep learning method for phage lifestyle classification~\cite{miao2024deephafier}. Based on a multilayer self-attention neural network combining protein information, DeePhafier directly extracts high-level features from a sequence by combining global self-attention and local attention and combines the protein features from genes to improve the performance of phage lifestyle classification. It has been shown that DeePhafier outperforms DeePhage and PhagePred. It would be interesting to compare the performance of DeePhafier and PhaTYP. 

\textbf{Phage virion protein annotation.} Phage virion proteins (PVPs) determine many biological properties of phages. In particular, they are effective at recognizing and binding to their host cell receptors without having deleterious effects on human or animal cells~\cite{kabir2022large}. Due to the very time-consuming and labor-intensive nature of experimental methods, PVP annotation remains a big challenge, which affects various areas of viral research, including viral phylogenetic analysis, viral host identification, and antibacterial drug development. Various ML methods have been developed to solve the PVP annotation problem~\cite{kabir2022large}. Those methods can be roughly classified into three groups: (1) traditional machine learning-based methods (using NB: naive bayes, RF: random forest, SCM: scoring card matrix, or SVM: support vector machine); (2) ensemble-based methods (using multiple machine learning models or training datasets), and (3) deep learning-based methods. Representative deep learning-based PVP classification methods are PhANNs~\cite{cantu2020phanns}, VirionFinder~\cite{fang2021virionfinder}, DeePVP~\cite{fang2022deepvp}, PhaVIP~\cite{shang2023phavip}, ESM-PVP~\cite{li2023esm}, and a PLM-based classifier~\cite{flamholz2024large}. PhANNs used $k$-mer frequency encoding and 12 MLPs as the classifiers. Both VirionFinder and DeePVP used CNN as classifiers. In VirionFinder, each protein sequence is represented by a ``one-hot'' matrix and a biochemical property matrix, while DeePVP only used one-hot encoding to characterize the protein sequence.  PhaVIP adapted a novel image classifier, Vision Transformer (ViT)~\cite{dosovitskiy2020image,raghu2021vision}, to conduct PVP classification. In particular, PhaVIP employed the chaos game representation (CGR) to encode k-mer frequency of protein sequence into images, and then leveraged ViT to learn both local and global features from sequence ``images''. The self-attention mechanism in ViT helps PhaVIP learn the importance of different subimages and their associations for PVP classification. ESM-PVP integrated a large pre-trained protein language model (PLM), i.e., ESM-2~\cite{lin2023evolutionary}, and an MLP to perform PVP identification and classification. A similar approach was proposed in~\cite{flamholz2024large}, where various pretrained PLMs~\cite{robson2022prose,elnaggar2021prottrans,brandes2022proteinbert}) were used.

\textbf{Phage-host interaction prediction.} Phages can specifically recognize and kill bacteria, which leads to important applications in many fields. Screening suitable therapeutic phages that are capable of infecting pathogens from massive databases has been a principal step in phage therapy design. Experimental methods to identify phage-host interactions (PHIs) are time-consuming and expensive; using high-throughput computational methods to predict PHIs is therefore a potential substitute. There are two types of computational methods for PHI prediction. One is alignment-based. We explicitly align the viral and bacterial whole-genome sequences and acquire matched sequences to indicate PHI. The other is alignment-free. We compare nucleotide features and/or protein features extracted from viral and bacterial genomes, and predict PHI using machine learning~\cite{nie2024advances}. Each type of method has its pros and cons. For example, PHIAF~\cite{li2022phiaf} is a deep learning method based on date augmentation, feature fusion, and the attention mechanism.  It first applies a GAN-based data augmentation module, which generates pseudo-PHIs to alleviate the data scarcity issue. Then it fuses the features originating from DNA and protein sequences for better performance.  Finally, it incorporates an attention mechanism into CNN to consider different contributions of DNA/protein sequence-derived features, which provides interpretability of the predictions. GSPHI~\cite{swan2013prevalent} is a deep learning method for PHI prediction with complementing multiple information. It first initializes the node representations of phages and target bacterial hosts via a word embedding algorithm (word2vec). Then it uses a graph embedding algorithm (structural deep network embedding: SDNE) to extract local and global information from the interaction network. Finally, it uses an MLP with two hidden layers to detect PHIs. PHPGAT~\cite{liu2025phpgat} is a novel phage–host prediction model. It integrates a multimodal heterogeneous knowledge graph with an advanced Graph Attention Network framework (GATv2) to enhance phage-host interaction predictions. PHPGAT first constructs a knowledge graph by integrating phage–phage, host–host, and phage–host interactions, capturing complex relationships between biological entities. It then employs GATv2 to extract deep node features and learn dynamic interdependencies, generating context-aware embeddings. Finally, an inner product decoder computes the likelihood of interactions between phage and host pairs based on these embeddings.

Recently, a deep learning-based method SpikeHunter was developed to perform a large-scale characterization of phage receptor-binding proteins (i.e., tailspike proteins), which are essential for determining the host range of phages~\cite{yang2024large}. SpikeHunter uses the ESM-2 protein language model~\cite{lin2023evolutionary} to embed a protein sequence into a representative vector. Then it predicts the probability of that protein being a tailspike protein using a fully connected 3-layer neural network. A reference set of 1,912 tailspike protein sequences and 200,732 non-tailspike protein sequences was curated from the INPHARED database~\cite{cook2021infrastructure}. SpikeHunter identified 231,965 diverse tailspike proteins encoded by phages across 787,566 bacterial genomes from five virulent, antibiotic-resistant pathogens. Remarkably, 86.60\% (143,200) of these proteins demonstrated strong correlations with specific bacterial polysaccharides. The authors found that phages with identical tailspike proteins can infect various bacterial species that possess similar polysaccharide receptors, highlighting the essential role of tailspike proteins in determining host range. This work significantly enhances the understanding of phage specificity determinants at the strain level and provides a useful framework for guiding phage selection in therapeutic applications.

PHILM (Phage-Host Interaction Learning from Metagenomic profiles) is a approach developed to infer phage–host interactions (PHIs) directly from shotgun metagenomic data~\cite{yang2025advancing}. PHILM was trained to predict prokaryotic abundance profiles from phage abundance profiles, and then infers putative PHIs through a sensitivity analysis of the trained model. PHILM was evaluated in synthetic datasets using ecological models with known ground truth and curated paired phage–prokaryotic abundance profiles from real metagenomes. Across all evaluations, PHILM consistently outperformed traditional co-abundance methods in recovering true interactions. When applied to 7,016 healthy human gut metagenomes, PHILM identified 90\% more genus-level PHIs than the assembly-based approach and revealed ecologically meaningful phage–host modules in the healthy gut. Moreover, the latent representations learned by PHILM captured microbial succession patterns in longitudinal data and served as powerful features for disease classification, outperforming taxonomic abundance-based features across multiple cohorts.

{\bf Phage lysins mining.} 
Phage lysins are enzymes produced by bacteriophages to degrade bacterial cell walls, allowing newly replicated phages to burst out of the host cell~\cite{schmelcher2021bacteriophage}. These enzymes specifically target and break down peptidoglycan, a major component of bacterial cell walls, causing rapid bacterial cell lysis and death. Phage lysins have garnered interest as potential therapeutic agents, especially given the rise of antibiotic-resistant bacteria. Unlike traditional antibiotics, lysins have a unique mechanism of action and can target specific bacterial species, reducing the risk of off-target effects on beneficial microbiota. However, experimental lysin screening methods pose significant challenges due to heavy workload. 

Very recently, AI techniques have been applied to discover novel phage lysins~\cite{zhang2024discovery,fu2024deepminelys}. DeepLysin is a unified software package to employ AI for mining the vast genome reservoirs  for novel antibacterial phage lysins~\cite{zhang2024discovery}. DeepLysin consists of two modules: the lysin mining module and the antibacterial activity prediction module. The input of the lysin mining module is assembled contigs. This module consists of four steps: coding sequences (CDS) annotation, redundancy removal, domain alignment, and transmembrane protein removal. This module utilizes traditional blastP/protein sequence alignment-based methods to identify putative lysins. The second module estimates the antibacterial activity of the putative lysins identified by the first module. This module utilizes multiple AI techniques, such as Word2vec and an ensemble classifier that integrates five common classifiers to differentiate diverse and complex protein features. It ultimately applies Logistic Regression as a non-linear activation function to produce final activity predictions as scores ranging from 0 to 1, with higher scores indicating increased antibacterial activity. One limitation of DeepLysin is that four types of manually selected features (i.e., composition-based feature, binary profile-based feature, position-based feature, physiochemical based feature) need to be provided to the classifier. The feature selection procedure apparently heavily relies on domain knowledge.  

DeepMineLys is a deep learning method based on CNN to identify phage lysins from human microbiome datasets ~\cite{fu2024deepminelys}. DeepMineLys started from collecting phage protein sequences to build training and test datasets. These protein sequences were then processed using two distinct embedding methods (TAPE~\cite{rao2019evaluating} and PHY~\cite{vazquez2021mining}). Each of the two embeddings was fed into a CNN to learn sequence information and generate representations separately. The two representations of TAPE and PHY were then concatenated into a final representation and fed into a densely connected layer for the final prediction. DeepMineLys leverages existing methods for processing protein sequence features. To some extent, it alleviates the burden of manual feature selection.

\section{Outlook}
In this review article, we introduced the applications of AI techniques in various application scenarios in microbiology and microbiome research. There are some common challenges in those applications. Here we summarize those challenges and offer tentative solutions to inform future research. 

\phantomsection
\subsection{Tradeoff between interpretability and complexity}
Machine learning models, especially deep learning models, often suffer from high complexity and low interpretability, hindering their application in clinical decision-making. Those models are often referred to as  black-box models, whose internal workings are hidden or not easily interpretable. In addition, deep learning models typically have more than thousands of neural weights whose training requires large sample sizes and high computational resources. We anticipate that those deep learning models can reach better performance than traditional machine learning models as long as the sample size is enough. However, in most clinic-related studies, traditional models (e.g., Random Forest) are still widely used due to their ease of implementation, smaller sample size requirement, and better interpretability. 

To address the interpretability issue, two different approaches can be employed. One approach is to employ methods such as SHAP (SHapley Additive exPlanations)~\cite{lundberg2017unified}, LIME (Local Interpretable Model-agnostic Explanations)~\cite{ribeiro2016should} to enhance the interpretability of black-box machine learning models. SHAP is a game-theoretic method used to explain the output of any machine learning model. It links optimal credit allocation to local explanations by leveraging Shapley values from game theory and their related extensions. LIME is a technique that approximates any black box machine learning model with a local, interpretable model to explain each individual prediction. By applying SHAP and LIME, we can gain insights into complex deep learning models, identify biases, and improve transparency, crucial for applications in microbiome research. 

The other approach is to employ ``white-box'' models, (whose internal logic, parameters, and decision process are fully transparent and human-readable). For instance, ReduNet~\cite{chan2022redunet} is a white-box deep network based on the principle of maximizing rate reduction. The authors argued that, at least in classification tasks, a key objective for a deep network is to learn a low-dimensional, linearly discriminative representation of the data. The effectiveness of this representation can be assessed by a principled measure from (lossy) data compression, i.e., rate reduction. Appropriately structured deep networks can then be naturally interpreted as optimization schemes designed to maximize this measure. The resulting multi-layer deep network shares key characteristics with modern deep learning architectures, but each component of ReduNet has a well-defined optimization, statistical, and geometric interpretation.  Applying ReduNet to microbiome data would be an interesting attempt. Unlike ReduNet, MDITRE is a supervised deep learning method specifically designed for microbiome research. It takes a phylogenetic tree, microbiome time-series data, and host status labels to learn human-interpretable rules for predicting host status~\cite{maringanti2022mditre}. The model consists of five hidden layers that can be directly interpreted in terms of if-then rule statements. The first layer focuses on phylogenetic relationships by selecting taxa relevant to predicting host status. The second layer focuses on time by identifying relevant time windows for prediction. The following layers determine whether the data from selected taxa and time windows exceed specific learned thresholds, and subsequently combine these conditions to generate the final rules for prediction.

\phantomsection
\subsection{The ``Small n, Large p'' issue}
Similar to many other omics studies, statistical or machine learning methods for microbiome research typically face the “small n, large p” issue, i.e., the number of parameters or microbial features (p) is much larger than the sample size (n).  This issue may result in overfitting, models behaving unexpectedly, providing misleading results, or failing completely. There are several classical strategies to deal with the ``small n, large p'' issue, e.g., feature selection, projection methods, and regularization algorithms.

Feature selection involves selecting a subset of features to use as input to predictive models.  Although the selection of an optimal subset of features is an NP-hard problem~\cite{chen1997minimum}, many compromised feature selection methods have been proposed. Those methods are often grouped into filtering, wrapped, and embedded methods~\cite{stanczyk2015feature}. For instance, GRACES is a GCN-based feature selection method~\cite{chen2023graph}. It exploits latent relations between samples with various overfitting-reducing techniques to iteratively find a set of optimal features which gives rise to the greatest decreases in the optimization loss. It has been demonstrated that GRACES significantly outperforms other feature selection methods on both synthetic and real-world gene expression datasets. It would be interesting to apply GRACES to microbiome data analysis.

Projection methods generate lower-dimensional representations of data while preserving the original relationships between samples. These techniques are often employed for visualization but can also serve as data transformations to reduce the number of predictors. Examples include linear algebra methods like SVD, PCA, and PCoA, as well as manifold learning algorithms, such as t-SNE, commonly used for visualization.

In standard machine learning models, regularization can be introduced during training to penalize the use or weighting of multiple features, promoting models that both perform well and minimize the number of predictors. This acts as an automatic feature selection process, and can involve augmenting existing models (e.g., regularized linear and logistic regression) or employing specialized methods like LASSO or multivariate nonlinear regression~\cite{chakraborty2012bayesian}. Since no single regularization method is universally optimal, it’s advisable to conduct controlled experiments to evaluate various approaches.

Recently, it has been proposed to use promising deep learning techniques (e.g., transfer learning, self-supervised learning, semi-supervised learning, few-shot learning, zero-shot learning, etc.) to deal with the ``small n, large p'' issue~\cite{safonova2023ten}. For example, transfer learning involves pre-training a model on a large dataset and then fine-tuning it on a smaller, task-specific dataset~\cite{pan2009survey}. By leveraging knowledge from a related but larger dataset, the pre-trained model can transfer learned representations to the small dataset, helping mitigate the issue of insufficient data.  Self-supervised learning is an approach to creating supervisory signals from the data itself, eliminating the need for labeled data~\cite{geiping2023cookbook}. This approach can effectively learn useful representations even with limited labeled data, as the model can train on unlabeled data, which is usually more abundant. In microbiome research, self-supervised techniques can use metagenomics sequences without annotations to learn meaningful patterns, later applied to the small labeled subset. Semi-supervised learning leverages a small amount of labeled data and a large amount of unlabeled data to train the model. Since the labeled data is small (small n), semi-supervised learning helps by learning from both labeled and unlabeled data to improve generalization. Few-shot learning enables models to generalize from very few examples~\cite{wang2020generalizing}. Few-shot learning techniques are specifically designed to handle scenarios with limited training data. They can quickly adapt to new tasks with only a handful of training samples. In personalized medicine, few-shot learning can help tailor models to individual patient data even when there is limited patient-specific training data. Zero-shot learning enables models to make predictions for classes they have not been explicitly trained on by learning from related classes or tasks~\cite{xian2018zero}.  This approach is especially useful when the data for certain categories or conditions is entirely missing (n = 0), allowing models to generalize from related categories or contexts. Deep learning models, especially those trained using self-supervised and transfer learning methods, can handle the high-dimensional feature space (large p) because they are adept at extracting useful features or representations from complex data. These approaches mitigate the problem of small sample sizes by either leveraging external data (e.g., transfer learning) or creating more efficient learning algorithms (e.g., few-shot and zero-shot learning). Applying those promising deep learning techniques to microbiome research to deal with the ``small n, large p'' issue would be very interesting. Some of the deep learning methods (especially those methods based on LLMs) discussed in this Review have already leveraged some of those techniques (e.g., transfer learning).

\phantomsection
\subsection{Benchmarking evaluations}
As we mentioned in previous sections several times, benchmarking evaluations are typically lacking in microbiology and microbiome research. Currently, there is no standardized pipeline for benchmarking machine learning or deep learning methods in microbiology and microbiome research. To ensure reproducibility across studies, it’s critical to standardize data preprocessing, which includes consistent methods for data collection, bioinformatics pipelines, and the profiling of microbiome taxonomies. Additionally, if feature dimension reduction is needed, it must be unbiased, using standardized methods for feature selection or reduction that apply uniformly across studies. Importantly, feature engineering should only be applied to training data and later evaluated on test data to avoid data leakage or overfitting. Furthermore, the creation of publicly available, well-annotated benchmarking datasets (analogous to MNIST or ImageNet in computer science) would provide the microbiome research community with reliable tools to assess and compare different machine learning models. Such datasets would accelerate progress and provide a framework for objective evaluation of new computational methods. Some attempts have been made in this regard. For example, MicrobiomeHD is a standardized database that compiles human gut microbiome studies related to health and disease~\cite{duvallet2017meta}. It contains publicly available 16S data from published case-control studies, along with associated patient metadata. The raw sequencing data for each study were obtained and processed using a standardized pipeline. The curatedMetagenomicData package is another excellent example of benchmark microbiome datasets. It offers uniformly processed human microbiome data, including bacterial, fungal, archaeal, and viral taxonomic abundances, as well as quantitative metabolic functional profiles and standardized participant metadata~\cite{pasolli2017accessible}. This comprehensive, curated collection of metagenomic data is well-documented and easily accessible, making it suitable for benchmarking machine learning methods.

Establishing benchmark datasets is critical for advancing AI application in microbiology and microbiome research. Such datasets can serve as standardized benchmarks that enable consistent, unbiased comparisons of algorithms, promote the development of robust predictive models, and foster reproducibility and transparency by allowing the research community to evaluate AI methods on a level playing field. Similar to the successful DREAM challenges in genomics, a community-driven effort to create public benchmarking datasets will foster collaboration, accelerate discovery, and establish best practices for AI approaches in microbiology and microbiome research. Collaborative input is vital for making this a reality. 

\phantomsection
\subsection{Recent Breakthroughs}
The rapid acceleration of AI development is widely recognized and driven by several converging factors—exponential increases in computing power, major global investments, and fundamental breakthroughs in algorithmic architectures. This unprecedented progress is reshaping scientific disciplines, and microbiology is no exception. We believe that microbiology and microbiome research stand to benefit profoundly from the expanding capabilities of modern AI systems. Below, we highlight two examples—AlphaFold 3 and Evo—that exemplify how generative and foundation models are beginning to transform our understanding of microbial life from sequence to ecosystem scale.

AlphaFold 3~\cite{abramson2024accurate} builds on the groundbreaking achievements of AlphaFold 2—recognized by the 2024 Nobel Prize for its impact on protein structure prediction—by extending its reach beyond individual proteins to entire biomolecular assemblies. AlphaFold 3 predicts 3D structures and interactions among proteins, DNA, RNA, small molecules, ions, and modified residues, effectively modeling ``all of life’s molecules” at atomic resolution. It employs a streamlined generative pipeline: the Pairformer module encodes pairwise relationships between atoms across biomolecules using efficient attention mechanisms; these representations are then refined by a diffusion model that transforms noisy atomic coordinates into physically plausible 3D structures. A neural network backbone with triangular attention integrates evolutionary and geometric constraints, enabling end-to-end modeling of complex molecular interactions with remarkable speed and accuracy.

For microbiome science, AlphaFold 3 is poised to revolutionize our ability to model microbial communities at the molecular level. By predicting not only protein structures but also networks of protein–protein, protein–metabolite, enzyme–substrate, and host–microbe interactions, it provides a framework for translating metagenomic sequences into testable, mechanistic hypotheses. This capability may enable: (1) decoding functional pathways in unculturable bacteria, revealing novel enzymes for bioremediation or antibiotic biosynthesis; (2) mapping microbial drug targets at atomic precision, accelerating discovery of probiotics and phage therapies; (3) simulating microbe–host crosstalk by predicting how microbial ligands modulate immune receptors (e.g., TLRs, GPCRs); and (4) engineering synthetic microbial consortia based on protein-level compatibility and metabolic exchange. By bridging sequence data and structural function, AlphaFold 3 has the potential to shift microbiome research from correlative to causal, molecular-level understanding—unlocking precision manipulation of microbial ecosystems across health, agriculture, and environmental applications.

Evo~\cite{nguyen2024sequence} represents a complementary advance-a genomic foundation model trained on massive genomic corpora, including 2.7 million prokaryotic and phage genomes ($\approx 300$ billion nucleotide tokens) and later expanded to 9.3 trillion DNA base pairs from a highly curated genome atlas spanning all domains of life in Evo 2~\cite{brixi2025genome}. It introduces a novel StripedHyena architecture that combines convolutional operators with rotary positional embeddings, allowing efficient long-context processing (up to one million nucleotides) with near-linear compute scaling. Trained via next-token prediction at single-nucleotide resolution, Evo learns the unified flow of biological information from DNA to RNA to protein—capturing motifs, regulatory grammar, and higher-order genome organization without task-specific fine-tuning.

In microbiology, Evo opens new avenues for predictive and generative genomics. It enables zero-shot identification of functional elements (e.g., promoters, splice sites, regulatory motifs), prediction of mutation effects, and de novo design of biological sequences. Applied to microbial systems, Evo could accelerate the design of synthetic microbes, precision-engineered phage cocktails, and enzymes for antibiotic resistance mitigation. For microbiome research, Evo’s multiscale modeling facilitates causal inference in community dynamics—linking genetic variation in unculturable gut bacteria to metabolic outputs, immune modulation (e.g., the interaction between short-chain fatty acids and toll-like receptors), or interspecies communication such as quorum sensing. When integrated with CRISPR-based editing and DNA synthesis platforms, Evo may ultimately enable programmable microbiome therapeutics and synthetic ecosystems for human and planetary health. As these capabilities expand, they also underscore the need for thoughtful consideration of biosafety and ethical governance in the era of AI-driven biological design.

\section{Acknowledgments}
We thank Yiyan Yang and Dan Huang for valuable suggestions and for carefully examining the manuscript. Y.-Y.L. acknowledges funding support from the National Institutes of Health (R01AI141529, R01HD093761, RF1AG067744, UH3OD023268, U19AI095219 and U01HL089856) as well as the Office of the Assistant Secretary of Defense for Health Affairs, through the Traumatic Brain Injury and Psychological Health Research Program (Focused Program Award) under award no. W81XWH-22-S-TBIPH2, endorsed by the Department of Defense. X.-W.W. acknowledges funding support from the National Institutes of Health (K25HL166208).

\section{Declaration of interests}
The authors declare no competing interests.

\bibliographystyle{unsrt}  
\bibliography{references}

\end{document}